\documentclass{vldb}

\usepackage {enumerate}
\usepackage [shortlabels]{enumitem}
\usepackage{graphicx}
\usepackage[square,comma,numbers,sort]{natbib}
\bibpunct{[}{]}{,}{n}{,}{,}
\usepackage{algorithm}
\usepackage{algorithmic}
\usepackage[caption=false]{subfig}
\usepackage{verbatimbox}
\usepackage{url}
\usepackage{times}
\usepackage{soul}
\usepackage{balance}  
\newdef{definition}{Definition}
\newdef{example}{Example}

\usepackage{color}
\newcommand{\edit}[1]{#1}
\newcommand{\editt}[1]{#1}
\newcommand{\TurboISO}{\textsf{Turbo}$_{\mathsf{ISO}}$}
\newcommand{\TurboISOp}{\textsf{Turbo}$_{\mathsf{HOM}}$}
\newcommand{\TurboISOP}{\textsf{Turbo}$_{\mathsf{HOM}}$\textsf{++}}

\begin{document}


\title{Taming Subgraph Isomorphism for RDF Query Processing}



%
%
%
%

\newcommand{\superscript}[1]{\ensuremath{^{\textrm{#1}}}}
\def\sharedaffiliation{\end{tabular}\newline\begin{tabular}{c}}

\def\ao{\superscript{\#}}
\def\ap{\superscript{\dag}}

\numberofauthors{4} 

\author{
\alignauthor
Jinha Kim \ap\ao \\
\email{jinha.kim@oracle.com}
\alignauthor
Hyungyu Shin \ap \\
\email{hgshin@dblab.postech.ac.kr}
\alignauthor
Wook-Shin Han \ap \thanks{corresponding author}\\
\email{wshan@postech.ac.kr}
\and
\auwidth=\textwidth
\alignauthorlow
\makebox{Sungpack Hong \ao \hspace*{0.3in} Hassan Chafi \ao \hspace*{0.6in}}\\
\email{\{sungpack.hong, hassan.chafi\}@oracle.com \hspace*{0.4in}}
\sharedaffiliation
\begin{tabular}{ccc}
    \affaddr{{\ap}POSTECH, South Korea{\ }} & & \affaddr{{\ao}Oracle Labs, USA{\ }} \\
\end{tabular}
}

\maketitle

\begin{abstract}
  RDF data are used to model knowledge in various areas such as life
  sciences, Semantic Web, bioinformatics, and social graphs. The size
  of real RDF data reaches billions of triples. This calls for a
  framework for efficiently processing RDF data. The core function of
  processing RDF data is subgraph pattern matching. There have been
  two completely different directions for supporting efficient
  subgraph pattern matching. One direction is to develop specialized
  RDF query processing engines exploiting the properties of RDF data
  for the last decade, while the other direction is to develop
  efficient subgraph isomorphism algorithms for general, labeled
  graphs for over 30 years. Although both directions have a similar
  goal (i.e., finding subgraphs in data graphs for a given query
  graph), they have been independently researched without clear
  reason. We argue that a subgraph isomorphism algorithm can be easily
  modified to handle the graph homomorphism, which is the RDF pattern
  matching semantics, by just removing the injectivity constraint. In
  this paper, based on the state-of-the-art subgraph isomorphism
  algorithm, we propose \edit{an in-memory solution,}
  \TurboISOP{}\edit{,} which is tamed for the RDF processing, and
  we compare it with the representative RDF processing engines for
  several RDF benchmarks in a server machine where billions of triples
  can be loaded in memory.
  In order to speed up
  \TurboISOP{}, 
  we also provide a simple yet effective transformation and a series
  of optimization techniques. Extensive experiments using several RDF
  benchmarks show that
  \TurboISOP{} 
  \emph{consistently and significantly} outperforms the representative
  RDF engines. Specifically, \TurboISOP{} outperforms its competitors
  by up to \emph{five} orders of magnitude.
\end{abstract}

\section{Introduction}
\label{sec:introduction}

The Resource Description Framework (RDF) is a standard for representing knowledge on the web. It is primarily designed for building the Semantic web and has been widely adopted in database and data mining communities. RDF models a fact as a triple which consists of a subject (S), a predicate (P), and an object (O). Due to its simple structure, many practitioners materialize their data in an RDF format. For example, RDF datasets are now pervasive in various areas including life sciences, bioinformatics, and social networks.
The size of real RDF data reaches billions of triples. Such billion-scale RDF data are fully loaded in main memory of today's server machine (The cost of a 1TB machine is less than \$40,000).


The SPARQL query language is a standard language for querying RDF data
in a declarative fashion. Its core function is subgraph pattern
matching, which corresponds to finding all graph homomorphisms in the
data graph for a query graph \cite{NW:10aa}.

In recent years, there have been significant efforts to speed up the processing of SPARQL queries by developing novel RDF query processing engines. Many engines \cite{NW:10aa,NW:10ab,YLW:13aa,WW:06aa,WKB:08aa,AMM:09aa} model RDF data as tabular structures and process SPARQL queries using specialized join methods. For example, RDF-3X \cite{NW:10aa} treats RDF data as an edge table, \textsc{Edge(S,P,O)}, and materializes six different orderings for this table, so that it can support many SPARQL queries just by using merge based join. Note that this approach is efficient for both disk-based and in-memory environments since merge join exploits only sequential scans. Some engines \cite{ACZ:10aa,ZYW:13aa,ZMC:11aa} treat RDF data as graphs (or matrices) and develop specialized graph processing methods for processing SPARQL queries. For example, gStore \cite{ZMC:11aa} uses specialized index structures to process SPARQL queries. Note that these index structures are based on gCode \cite{ZCY:08aa}, which was originally proposed for graph indexing. 

Subgraph isomorphism, on the other hand, has been studied since the 1970s. The representative algorithms are VF2 \cite{PFS:04aa}, QuickSI \cite{SZL:08aa}, GraphQL \cite{HS:08aa}, GADDI \cite{ZLY:09aa}, SPATH \cite{ZH:10aa}, and \TurboISO{} \cite{HLL:13aa}. In order to speed up performance, these algorithms exploit good matching orders and effective pruning rules. A recent study \cite{LHK:12aa} shows that good subgraph isomorphism algorithms significantly outperform graph indexing based ones. However, all of these algorithms use only small graphs in their experiments, and thus, it still remains unclear whether these algorithms can show good performance for billion-scale graphs such as RDF data.

Although subgraph isomorphism processing and RDF query processing have
similar goals (i.e., finding subgraphs in data graphs for a given
query graph), they have two inexplicably different directions. A
subgraph isomorphism algorithm can be easily modified to handle the
graph homomorphism, which is the RDF pattern matching semantics, just
by removing the injectivity constraint.

In this paper, based on the state-of-the-art subgraph isomorphism
algorithm \cite{HLL:13aa}, we propose \edit{an in-memory solution,}
\TurboISOP{}\edit{,} which is \edit{tamed} for the RDF processing, and
we compare it with the representative RDF processing engines for
several RDF benchmarks in a server machine where billions of triples
can be loaded in memory.
We believe that this
approach opens a new direction for RDF processing so that both
traditional directions can merge or benefit from each other.

By transforming RDF graphs into labeled graphs, we can apply
  subgraph homomorphism methods to RDF query processing. Extensive
  experiments using several benchmarks show that a direct modification of
  \TurboISO{} outperforms the RDF processing engines for queries which
  require a small amount of graph exploration. However, for some
  queries which require a large amount of graph exploration, the direct
  modification is slower than some of its competitors. This poses an important research question: ``\emph{Is this phenomenon due to inherent limitations of the graph homomorphism (subgraph isomorphism) algorithm?}''
  Our profile results show that two major subtasks
  of \TurboISO{} --- 1) exploring candidate subgraphs in $ExploreCandidateRegion$ and 2)
  enumerating solutions based on candidate regions in $SubgraphSearch$ ---
 require performance improvement. \TurboISOP{} resolves such performance
 hurdles by proposing the type-aware transformation and tailored
 optimization techniques.



\edit{First}, in order to speed up $ExploreCandidateRegion$, we propose a novel
  transformation (Section~\ref{sec:type-awre-transf}), called
\emph{type-aware transformation}, which is simple yet effective in
processing SPARQL queries. In type-aware transformation, by embedding
the types of an entity (i.e., a subject or object) into a vertex label
set, we can eliminate corresponding query vertices/edges from a query
graph. With type-aware transformation, the query graph size decreases,
its topology becomes simpler than the original query, and thus, this
transformation improves performance accordingly by reducing the amount
of graph exploration.

In order to optimize performance in depth, in both $Explore\-Can\-didate\-Region$ and $Subgraph\-Search$, we propose a series of optimization techniques (Section~\ref{sec:optimization}), each of which contributes to performance
improvement significantly for such slow queries. In addition, we explain how
\TurboISOP{}  is extended to support  1) general SPARQL
  features such as OPTIONAL, and FILTER, and 2) parallel execution
 for \TurboISOP{} in \edit{a non-uniform memory access (NUMA)}
 architecture \edit{\cite{LPM:13aa,LBK:14aa}}.
These general features are necessary to execute comprehensive
benchmarks such as \edit{Berlin SPARQL benchmark (BSBM) \cite{BS:09aa}}.
Note also that, when the RDF data size grows large, we have to rely on the NUMA
architecture.

Extensive experiments using several representative benchmarks show that
\TurboISOP{} consistently and significantly outperforms all its
competitors for all queries tested. Specifically, our method
outperforms the competitors by up to five orders of magnitude with
only a single thread. This indicates that a subgraph isomorphism
algorithm tamed for RDF processing can serve as an in-memory RDF
accelerator on top of a commercial RDF engine for \emph{real-time} RDF
query processing.

Our contributions are as follows. 1) We provide the first direct
comparison between RDF engines and the state-of-the-art subgraph
isomorphism method tamed for RDF processing, \TurboISOP{}, thro-ugh extensive experiments and analyze experimental results in
depth. 2) In order to simplify a query graph, we propose a
novel transformation method called type-aware transformation, which
contributes to boosting query performance. 3) In order to speed up
query performance further, we propose a series of performance
optimizations as well as NUMA-aware parallelism for fast RDF query
processing. 4) Extensive experiments using several benchmarks show
that the optimized subgraph isomorphism method consistently and
significantly outperforms representative RDF query processing engines.

The rest of the paper is organized as follows.
Section~\ref{sec:preliminary} describes the subgraph isomorphism, its
state-of-the-art algorithms, \TurboISO{}, and their modification for
the graph homomorphism.  Section~\ref{sec:sparql-proc-subgr} presents
how a direct modification of \TurboISO{}, \TurboISOp{}, handles the
SPARQL pattern matching. Section~\ref{sec:turbohom++} describes how we
obtain \TurboISOP{} from \TurboISOp{} using the type-aware
transformation and optimizations for the efficient
SPARQL pattern matching.
\iftr
\edit{Section~\ref{sec:extension} describes
  how \TurboISOP{} can handle general SPARQL features and discusses
  the parallelization of \TurboISOP{}}.
\fi
Section~\ref{sec:related-work}
reviews the related work.  Section~\ref{sec:experiment} presents the
experimental result.  Finally, Section~\ref{sec:conclusion} presents
our conclusion.
\iftr \else Note that due to the space limit, please refer \cite{tr}
for how \TurboISOP{} handle OPTIONAL, UNION, FILTER keywords, and
parallelize.  \fi

\section{Preliminary}
\label{sec:preliminary}


\subsection{Subgraph Isomorphism and RDF Pattern Matching Semantic}
\label{sec:subgraph-isomorphism}

Suppose that a labeled graph is defined as $g(V,E,L)$, where $V$ is a
set of vertices, $E (\subseteq V \times V)$ is a set of edges, and $L$
is a labeling function which maps from a vertex or an edge to the
corresponding label set or label, respectively. Then, the subgraph
isomorphism is defined as follows.

\begin{definition}
  \label{def:subiso}
  \cite{LHK:12aa} Given a query graph $q(V,E,L)$ and a data graph
  $g(V',E',L')$, a \textit{subgraph isomorphism} is an injective
  function $M: V \to V'$ such that 1) $\forall v \in V, L(v) \subseteq
  L'(M(v))$ and 2) $\forall (u,v) \in E, (M(u),M(v)) \in E'$ and
  $L(u,v) = L'(M(u), M(v))$.
\end{definition}

If a query vertex, $u$, has a blank label set (or does not specify vertex label equivalently), it can match any data vertex.
Here, $L(u)=\emptyset$, and thus, the subset condition, $L(u) \subseteq L'(M(u))$, is always satisfied.
Similarly, if a query edge $(u,v)$ has a blank label, it can match any data edge by generalizing the \edit{equality} condition
$L(u,v) = L'(M(u), M(v))$ to $L(u,v) \subseteq L'(M(u), M(v))$.

The graph homomorphism \cite{FLM:10aa} is easily obtained from the
subgraph isomorphism by just removing the injective constraint on $M$
in Definition~\ref{def:subiso}. Even though the RDF pattern matching
semantics is based on the graph homomorphism, to answer SPARQL queries
which have variables on predicates, a mapping from a query edge to an
edge label is also required.  We call such graph homomorphism the
\emph{e(xtended)-graph homomorphism} and present a formal definition
for it as follows.

\begin{definition}
  \label{def:subhomo}
  Given a query graph $q(V,E,L)$ and a data graph $g(V',E',L')$, an
  \textit{e(xtended)-graph homomorphism} is a pair of two mapping
  functions, a query vertex to data vertex function $M_{v}: V \to V'$
  such that 1) $\forall v \in V, L(v) \subseteq L'(M_v(v))$ and 2)
  $\forall (u,v) \in E, (M_v(u),M_v(v)) \in E'$, and $L(u,v) =
  L'(M_v(u), M_v(v))$, and a query edge to edge label function $M_{e}:
  V \times V \to L$ such that $\forall (u,v) \in E, M_e(u,v) =
  L'(M_v(u), M_v(v)).$
\end{definition}

The subgraph isomorphism problem (resp. the e-graph homomorphism
problem) is to find all distinct subgraph isomorphisms (resp. e-graph
homomorphisms) of a query graph in a data graph.


Figure~\ref{fig:iso-example-2} shows a query $q_1$ and a data graph
$g_1$. In $q_1$, \verb|_| means a blank vertex label set or blank edge
label. In the subgraph isomorphism, there is only one solution --
$M^1 =\{(u_0,v_0), $ $(u_1,v_{1}), (u_2,$
$v_{2}), (u_3, v_3), (u_4, v_4)\}$. In the e-graph homomorphism, there
are three solutions -- $M_{v}^1 = M^1$,
$M_e^1 = \{((u_0,u_1),a),((u_0,u_4),b),$
$((u_2,u_1),a),((u_2,u_3),a),((u_3,u_4),c) \}$, $M_{v}^2 = $
$\{(u_0,v_2),$ $(u_1, $ $v_3), (u_2,v_2),(u_3,v_3),(u_4,v_5)\}$,
$M_e^2 = M_e^1$, and $M_{v}^3 =\{(u_0,v_2), $
$(u_1,v_1),(u_2,v_2),(u_3,v_3),(u_4,v_5)\}$, $M_e^3 = M_e^1$.

\begin{figure}[h!]
  \centering
  \subfloat[query graph $q_1$.]
  {
    \label{fig:iso-query-2}
    \includegraphics[width=0.35\linewidth]{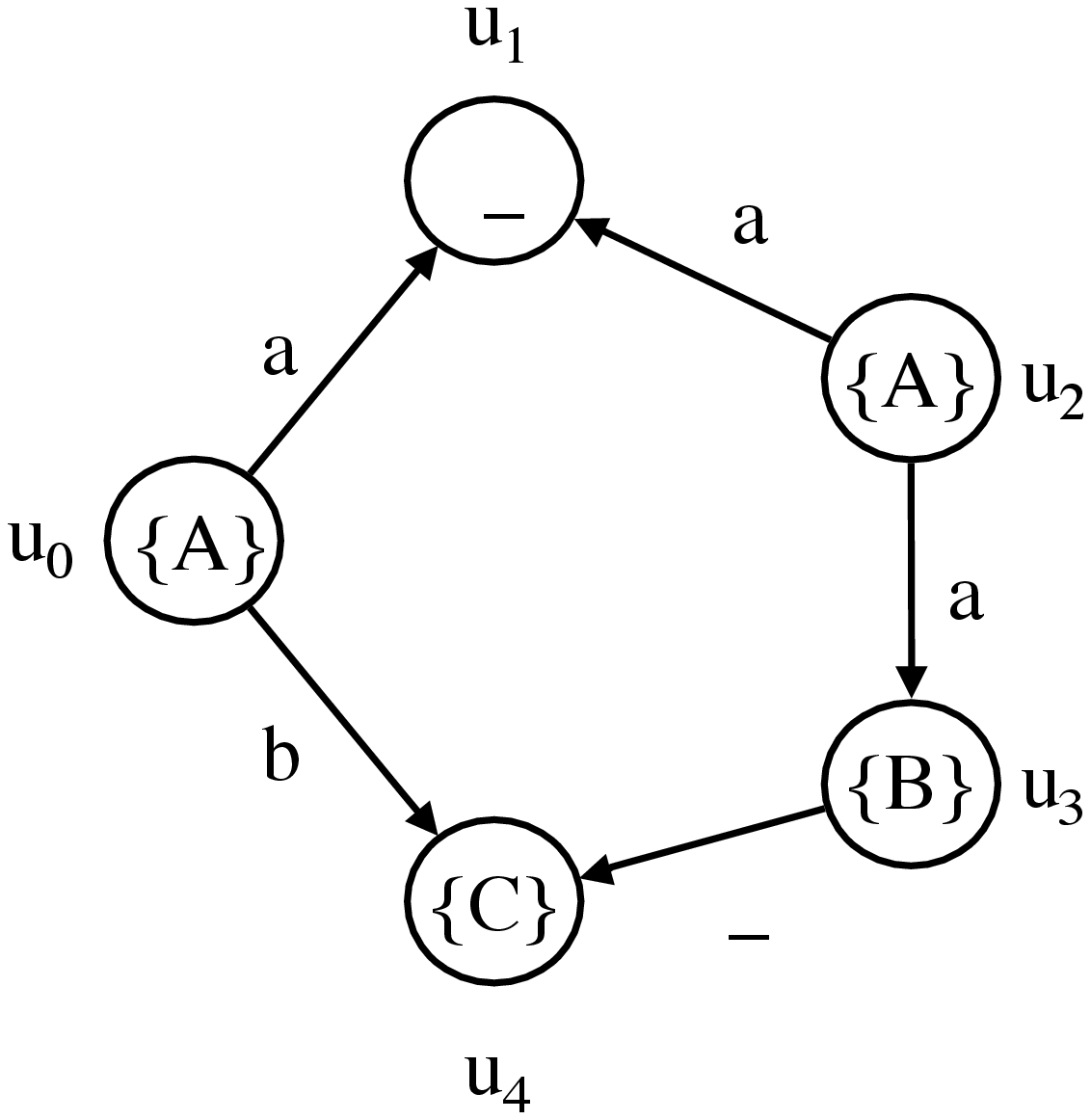}
  }
  \subfloat[data graph $g_1$.]
  {
    \label{fig:iso-data-2}
    \includegraphics[width=0.55\linewidth]{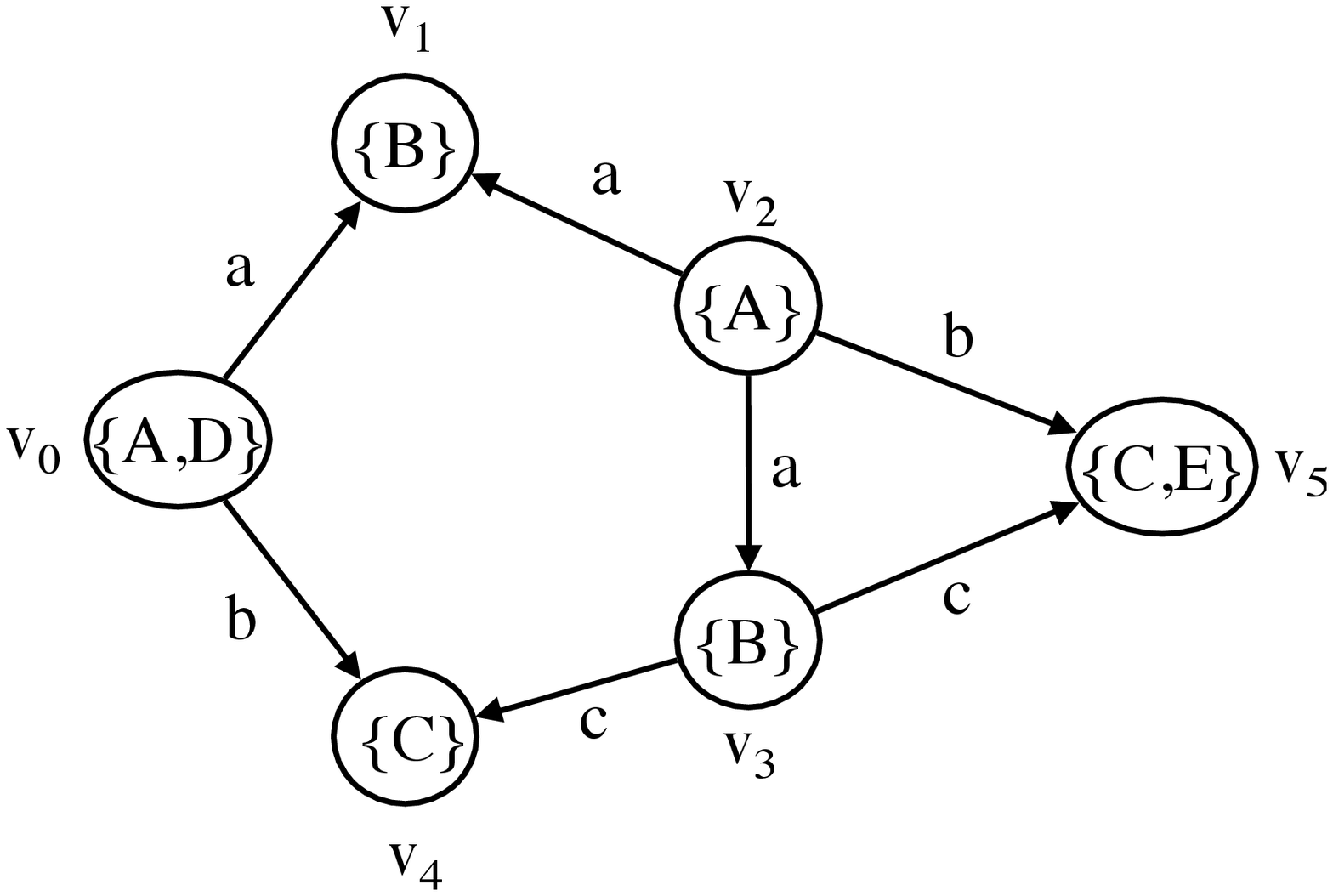}
  }
  \vspace{-2mm}
  \caption{Example of subgraph isomorphism and e-graph homomorphism.}
  \label{fig:iso-example-2}
  \vspace{-2mm}
\end{figure}

\subsection{Turbo$_{\text{ISO}}$}
\label{sec:turbo_iso}

In this subsection, we introduce the state-of-the art subgraph
isomorphism solution, \TurboISO{}\cite{HLL:13aa}, and its modification
for the e-graph homomorphism. Although we only describe the
modification of \TurboISO{} for the e-graph homomorphism, such
modification is applicable to other subgraph isomorphism algorithms
including VF2 \cite{PFS:04aa}, QuickSI \cite{SZL:08aa}, GraphQL
\cite{HS:08aa}, GADDI \cite{ZLY:09aa}, and SPATH \cite{ZH:10aa}, since
all of the subgraph algorithms mentioned are instances of a generic
subgraph isomorphism framework \cite{LHK:12aa}.

\TurboISO{} presents an effective method for the notorious
\emph{matching order} problem from which all the previous subgraph
isomorphism algorithms have suffered~\cite{LHK:12aa}.
Figure~\ref{fig:iso-example} illustrates an example of the matching
order problem, where $q_2$ is the query graph, and $g_2$ is the data
graph\footnote{For simplicity, we omit the edge labels and allow only
  one vertex label in the data graph.}. Note that this example query
results in no answers. However, the time to finish this query can
differ drastically by how one chooses the matching order, as it leads
to different number of comparisons. For instance, a matching order
$<u_0, u_2, u_1, u_3>$ requires $1 + 10000*10*5$ comparisons while a
different matching order $<u_0, u_3, u_1, u_2>$ requires only 1 + 5 *
10 comparisons.

\begin{figure}[h!]
  \vspace{-5mm}
  \centering
  \subfloat[query graph $q_2$.]
  {
    \label{fig:iso-query}
    \raisebox{13mm}
    {
      \includegraphics[width=0.3\linewidth]{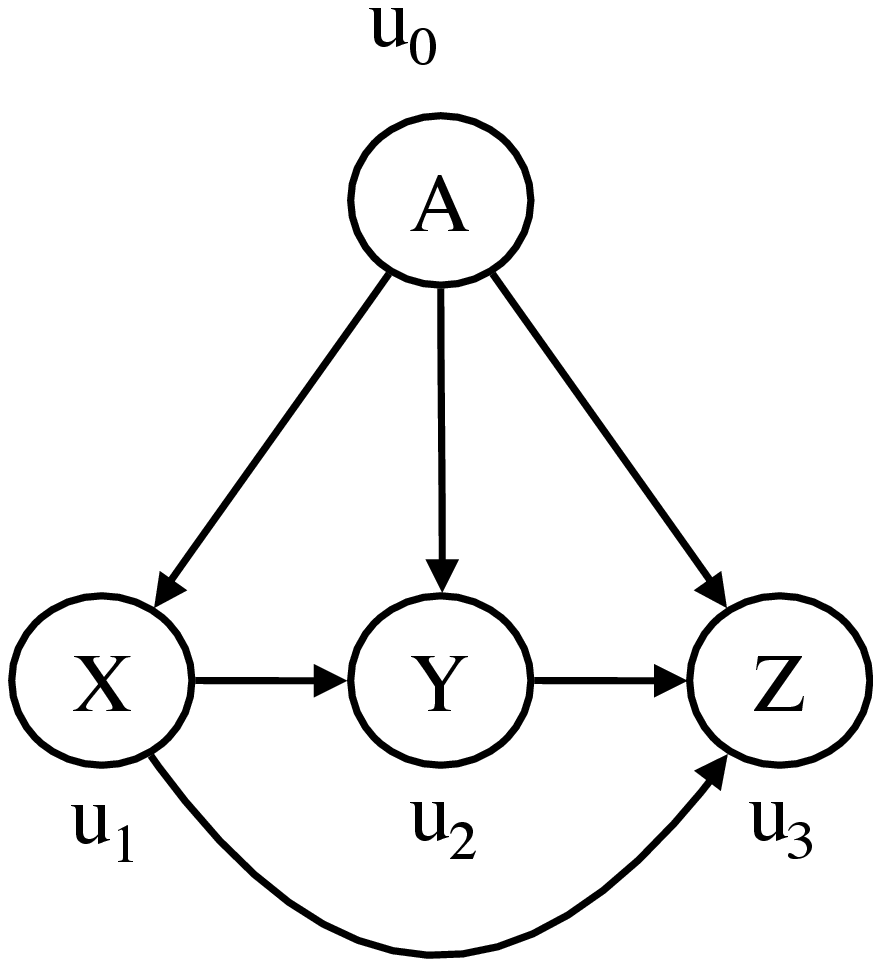}
    }
  }
  \subfloat[data graph $g_2$.]
  {
    \label{fig:iso-data}
    \includegraphics[width=0.7\linewidth]{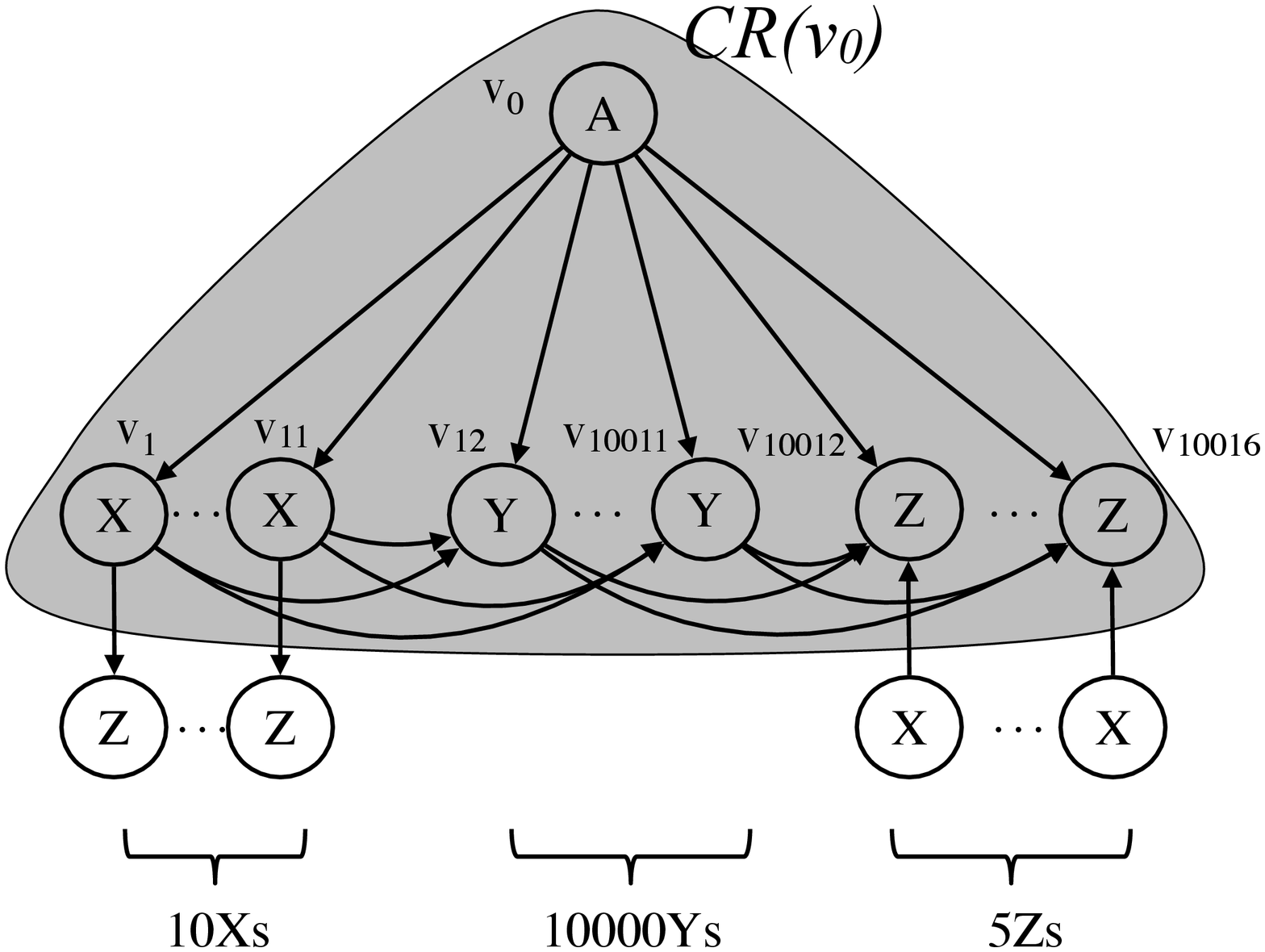}
  }
  \vspace{-1mm}
  \caption{Example of showing the matching order problem.}
  \label{fig:iso-example}
\end{figure}

\TurboISO{} solves the matching order problem with \emph{candidate region
exploration}, a technique that accurately estimates the number of candidate
vertices for a given query path \cite{HLL:13aa}. In particular, \TurboISO{} first
identifies candidate data subgraphs (i.e., candidate regions) from the starting vertices (e.g. the shaded area in
Figure~\ref{fig:iso-data}), then explores each region by performing a
depth-first search, which allows almost exact selectivity for each query path.

Algorithm~\ref{alg:turboiso} outlines the overall procedure of
\TurboISO{} in detail. First, if a query graph has only one vertex $u$
and no edge, it is sufficient to retrieve all data vertices which have
$u$'s labels ($=V(g)_{L(u)}$) and to find a subgraph isomorphism for
each of them (lines 2--4). Otherwise, it selects the starting query
vertex from the query graph (line 6). Then, it transforms the query
graph into its corresponding query tree (line 7). After getting the
query tree, for each data vertex that contains the vertex label of the
starting query vertex, the candidate region is obtained by exploring
the data graph (lines 9). If the candidate region is not empty, its
matching order is determined (line 11). The data vertex, $v_s$, is
mapped to the first query vertex $u_s$ by assigning $M(u_s) = v_s$ and
$F(v_s) = true$ where $F: V \to boolean$ is a function which checks
whether a data vertex is mapped or not (line 12). Then, the remaining
subgraph matching is conducted (line 13). Lastly, the mapping
$(u_s, v_s)$ is restored by removing the mapping for $u_s$ and
assigning $F(v_s) = false$ (line 14).

\begin{algorithm}[h!]
  \caption{\TurboISO{}($g(V,E,L),q(V',E',L')$)}
  \begin{algorithmic}[1]
    \REQUIRE $q$: query graph, $g$: data graph
    \ENSURE all subgraph \edit{isomorphisms} from $q$ to $g$.
    \IF {$V(q) = \{u\}$ and $E = \phi$}
      \FOR {\textbf{each } $v \in V(g)_{L(u)}$}
        \STATE report $M = \{(u,v)\}$
      \ENDFOR
    \ELSE
      \STATE $u_s \leftarrow ChooseStartQueryVertex(q,g)$
      \STATE $q' \leftarrow WriteQueryTree(q,u_s)$
      \FOR{\textbf{each} $v_s \in \{v|v \in V, L(u_s) \subseteq L(v) \}$}
        \STATE $CR \leftarrow ExploreCandidateRegion(u_s,v_s)$
        \IF{$CR$ is not empty}
          \STATE $order \leftarrow DetermineMatchingOrder(q',CR)$
          \STATE $UpdateState(M,F,u_s,v_s)$
          \STATE $SubgraphSearch(q,q',g,CR,order,1)$
          \STATE $RestoreState(M,F,u_s,v_s)$
        \ENDIF
      \ENDFOR
    \ENDIF
  \end{algorithmic}
  \label{alg:turboiso}
\end{algorithm}


\textbf{ChooseStartQueryVertex. }$Choose\-Start\-Query\-Vertex$ tries
to pick the starting query vertex which has the least number of
candidate regions. First, as a rough estimation, the query vertices
are ranked by their scores. The score of a query vertex $u$ is
$rank(u) = \frac{freq(g, L(u))}{deg(u)}$, where $freq(g, L(u))$ is the
number of data vertices that have $u$'s vertex labels. The score
function prefers lower frequencies and higher degrees. After obtaining
the top-k least-scored query vertices, the number of candidate regions
is more accurately estimated for each of them by using the degree
filter and the neighborhood label frequency (NLF) filter. The degree
filter qualifies the data vertices which have equal or higher degree
than their corresponding query vertices. The NLF filter qualifies the
data vertices which have equal or larger number of neighbors for all
distinct labels of the query vertex. In Figure~\ref{fig:iso-example},
for example, $u_0$ becomes the starting query vertex since it has the
least number of candidate regions (= 1).


\textbf{WriteQueryTree. }Next, $WriteQueryTree$ transforms the query graph to the query tree.
From the starting query vertex obtained by
$Choose\-Start\-Query\-Vertex$, a breath-first tree traversal is
conducted. Every non-tree edge $(u,v)$ of the query graph also is
recorded in the corresponding query tree. For example, when $u_0$ is
the starting query vertex, the non-tree edges of $q_2$'s query tree
are $(u_1, u_2)$,$(u_1, u_3)$, and $(u_2,u_3)$.




\textbf{ExploreCandidateRegion. }Using the query tree and the starting
query vertex, $Explore\-Candidate\-Region$ collects the candidate
regions. A candidate region is obtained by exploring the data graph
from the starting query vertex in a depth-first manner following the
topology of the query tree. During the exploration, the injectivity
constraint should be enforced. The shaded area of
Figure~\ref{fig:iso-data} is the candidate region $CR(v_0)$ based on
$q_2$'s query tree. Note that the candidate region expansion is
conducted only after the current data vertex satisfies the constraints
of the degree filter and the NLF filter.


\textbf{DetermineMatchingOrder. }After obtaining the candidate regions
for a starting data vertex, the matching order is determined for each
candidate region. Using the candidate region,
$Determine\-Matching\-Order$ can accurately estimate the number of
candidate vertices for each query path. Then, it orders all query
paths in the query tree by the number of candidate vertices. 
For example, from $CR(v_0)$, the ordered list of query paths is [$u_0.u_3$, $u_0.u_1$,  $u_0.u_2$].
Thus, we can easily see that $<u_0, u_3, u_1,
u_2>$ is the best matching order based on this ordered list.


\textbf{SubgraphSearch. }Exploiting the data structures obtained from the previous steps,
$SubgraphSearch$ (Algorithm~\ref{alg:subgraphsearch}) enumerates all
distinct subgraph isomorphisms. It first determines the current query
vertex $u$ from a given matching order $order$ (line 1). Then, it
obtains a set of data vertices, $C_R$ from a candidate region $CR$
(line 2).  $CR(u,v)$ represents the candidate vertices of a query
vertex $u$ which are the children of $v$ in $CR$, and $P(q',u)$
is the parent of $u$ in a query tree $q'$. For each candidate data
vertex $v$, if $v$ has already been mapped, the current
solution is rejected since it violates the injectivity constraint of
the subgraph isomorphism (lines 4--6). Next, by calling $IsJoinable$,
if the query vertex $u$ of the current data vertex $v$ has non-tree
edges, the existence of the corresponding edges are checked in the
data graph (line 7). For example, given $CR(v_0)$ and the matching
order $<u_0, u_3, u_1, u_2>$, when making the embedding for $u_1$, we
must check whether there is an edge from $M(u_1)$ to $M(u_3)$. If the
$IsJoinable$ test is passed, the mapping information is updated by
assigning $M(u) = v$ and $F(v) = true$ (line 8). After updating the
mapping, if all query vertices are mapped, a subgraph isomorphism $M$
is reported (lines 9--10). Otherwise, further subgraph search is
conducted (line 12). Finally, all changes done by $UpdateState$ are
restored (line 14).


\begin{algorithm}
  \caption{$SubgraphSearch(q, q', g, CR, order, d_c)$}
  \begin{algorithmic}[1]
    \STATE $u \leftarrow order[d_c]$
    \STATE $C_{R} \leftarrow CR(u, M(P(q',u)))$
    \FOR {\textbf{each} $v \in C_{R}$ such that $v$ is not yet matched}
      \IF {$F(v) = true$}
        \STATE \textbf{continue}
      \ENDIF
      \IF {$IsJoinable(q,g,M,u,v,\dots)$}
        \STATE $UpdateState (M,F,u,v)$
        \IF {$|M| = V(q)$}
          \STATE \textbf{report} $M$
        \ELSE
        \STATE $SubgraphSearch(q,q',g,CR,order,d_c+1)$
        \ENDIF
        \STATE $RestoreState(M,F,u,v)$
      \ENDIF
    \ENDFOR
  \end{algorithmic}
  \label{alg:subgraphsearch}
\end{algorithm}


\textbf{Modifying \TurboISO{} for e-Graph Homomorphism. }We first
explain how the generic subgraph isomorphism algorithm \cite{LHK:12aa}
can easily handle graph homomorphism. The generic subgraph isomorphism
algorithm is implemented as a backtrack algorithm, where we find
solutions by incrementing partial solutions or abandoning them when it
is determined that they cannot be completed. Here, given a query graph
$q$ and its matching order ($u_{\sigma(1)}$, $u_{\sigma(2)}$, ...,
$u_{\sigma(|V(q)|)}$), a solution is modeled as a vector $\vec{v}$ =
($M({u_{\sigma(1)}})$, $M({u_{\sigma(2)}})$, ...,
$M({u_{\sigma(|V(q)|)}})$) where each element in $\vec{v}$ is a data
vertex for the corresponding query vertex in the matching order.  At
each step in the backtrack algorithm, if a partial solution is given,
we extend it by adding every possible candidate data vertex at the
end.  Here, any candidate data vertex that does not satisfy the
following three conditions must be pruned.

\begin{enumerate}[1)]
\setlength{\itemsep}{0pt}
\item $\forall u_i \in V(q)$, $L(u_i) \subseteq L(M(u_i))$
\item $\forall (u_i, u_j) \in E(q)$, $(M(u_{i}), M(u_{j})) \in E(g)$ and $L(u_{i}, u_{j}) = L(M(u_{i}), M(u_{j}))$
\item $M(u_i) \neq M(u_j)$ if $u_i \neq u_j$
\end{enumerate}

Note that the third condition ensures the injective condition, guaranteeing
that no duplicate data vertex exists in each solution vector.
Thus, by just disabling the third condition, the generic subgraph
isomorphism algorithm finds all possible homomorphisms.

Now, we describe how to disable the third condition in \TurboISO{}, which
is an instance of the generic subgraph isomorphism algorithm.
\TurboISO{} uses pruning rules by applying filters in 
$Explore\-CandidateRegion$ and $SubgraphSearch$. First, the degree filter and the NLF filter should be modified since a data vertex can be mapped to
multiple query vertices. The degree filter qualifies data vertices
which have an equal number or more neighbors than distinct labels of
their corresponding query vertices. The NLF filter qualifies data
vertices which have at least one neighbor for all distinct labels of
their corresponding query vertices.
Second, lines 4--6 of $SubgraphSearch$ ensuring the third condition should be removed in
order to disable the injectivity test. As we see here, with minimal
modification to \TurboISO{}, it can easily support graph homomorphism.

In order to make \TurboISO{} handle the e-graph homomorphism, the query
edge to edge label mapping, $M_e$, should be additionally added in $SubgraphSearch$.
For this, $UpdateState$ assigns $M_e(P(q',u), u)$ = $L(M_v(P(q',u)), M_v(u))$
, and $RestoreState$ removes such mapping.
From here on, let us denote
\TurboISO{} modified for the e-graph homomorphism as \TurboISOp{}.

\section{RDF Query  Processing by e-Graph Homomorphism}
\label{sec:sparql-proc-subgr}

In this section, we discuss how RDF datasets can be naturally viewed
as graphs (Section~\ref{sec:rdf-as-graph}), and thus how an RDF
dataset can be directly transformed into a corresponding labeled graph
(Section~\ref{sec:direct-transf}). After such a transformation,
henceforth, the subgraph isomorphism algorithms modified for the
  e-graph homomorphism such as \TurboISOp{} can be applied for
processing SPARQL queries.


\subsection{RDF as Graph}
\label{sec:rdf-as-graph}

An RDF dataset is a collection of triples each of which consists of a
subject, a predicate, and an object.
By considering triples as directed edges, an RDF dataset naturally
becomes a directed graph: the subjects and the objects are vertices
while the predicates are edges. Figure~\ref{fig:graph} is a graph
representation of triples that captures \edit{type} relationships between
university organizations. Note that we use rectangles to represent
vertices in RDF graphs to distinguish them from the labeled graphs.

\begin{figure}[h!]
  \centering
  \includegraphics[width=1\linewidth]{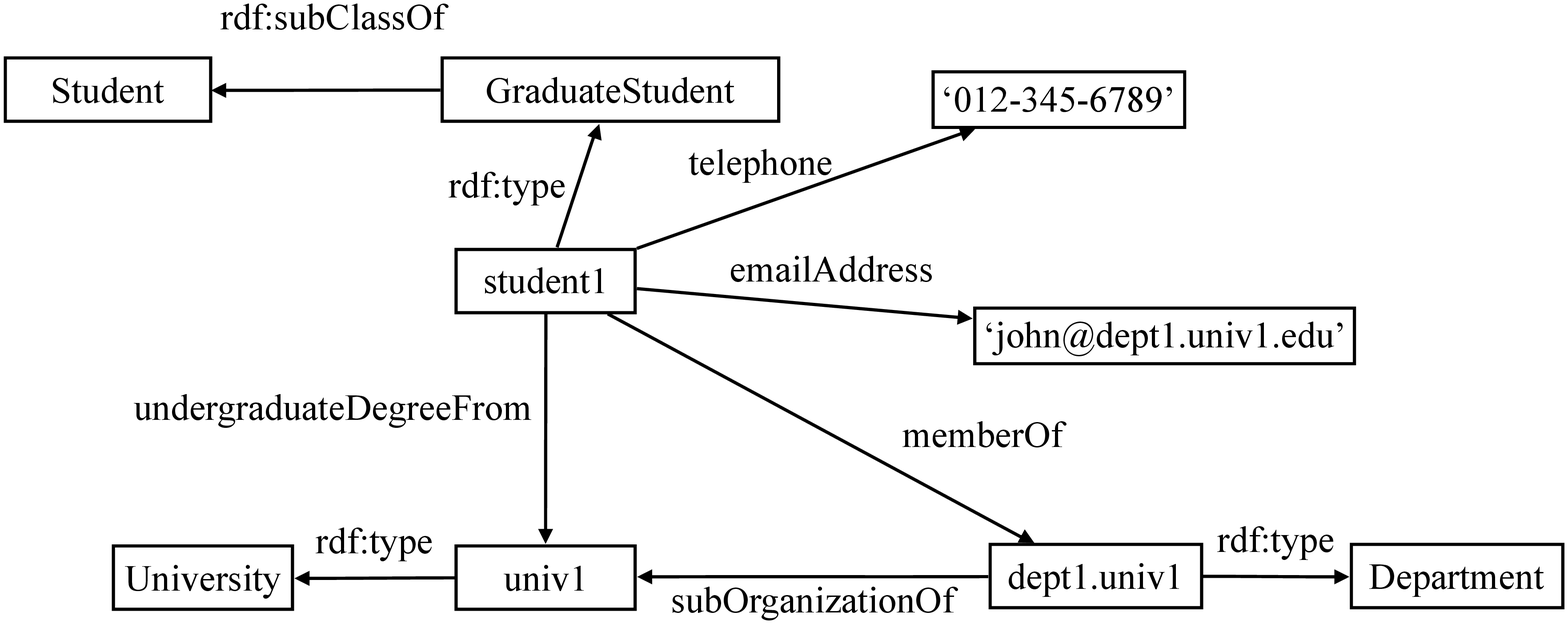}
  \vspace{-7mm}
  \caption{RDF graph.}
  \vspace{-2mm}
  \label{fig:graph}
\end{figure}

\subsection{Direct Transformation}
\label{sec:direct-transf}

To apply subgraph isomorphism algorithms modified for e-graph
homomorphism (e.g. \TurboISOp{}) for RDF query processing, RDF graphs
have to be transformed into labeled graphs first.




The most basic way to transform RDF graphs is (1) to map subjects and
objects to vertex IDs and (2) to map predicates to edge labels. We
call such transformation the \textit{direct transformation} because
the topology of the RDF graph is kept in the labeled graph after the
transformation. The vertex label function $L(v) (v \in V(g))$ is the
identity function (i.e. $L(v) = \{v\}$).

Figure~\ref{fig:direct-transformation} shows the result of the direct
transformation of Figure~\ref{fig:graph} --
Figures~\ref{fig:direct-sub-obj-map}, \ref{fig:direct-pred-map},
and~\ref{fig:direct-trans-graph} are the vertex mapping table , the
edge label mapping table, and the transformed graph, respectively.

\setlength{\tabcolsep}{0.5mm}
\begin{figure}[h!]
  \vspace{-2mm}
  \centering
  \subfloat[vertex mapping table.]
  {
    \label{fig:direct-sub-obj-map}
    \scriptsize
    \begin{tabular}[t]{|c|c|}
      \hline
      Subject/Object & Vertex \\
      \hline
	  GraduateStudent & $v_0$\\
	  Student  & $v_1$ \\
	  University  & $v_2$ \\
	  Department  & $v_3$ \\
	  student1  & $v_4$ \\
	  univ1  & $v_5$ \\
	  dept1.univ1  & $v_6$ \\
	  `012-345-6789' & $v_7$  \\
	  `john@dept1.univ1.edu' & $v_8$  \\
	  \hline
    \end{tabular}
  }
  \subfloat[edge label mapping table.]
  {
    \label{fig:direct-pred-map}
    \scriptsize
    \begin{tabular}[t]{|c|c|}
      \hline
      Predicate & Edge Label\\
      \hline
       rdf:type & a\\
       rdf:subClassOf & b\\
       undergradDegreeFrom & c\\
       memberOf & d\\
       subOrganizationOf & e\\
       telephone & f\\
       emailAddress & g\\
      \hline
    \end{tabular}
  }
  \\
  \vspace{-2mm}
  \subfloat[graph.]
  {
    \label{fig:direct-trans-graph}
    \includegraphics[width=0.6\linewidth]{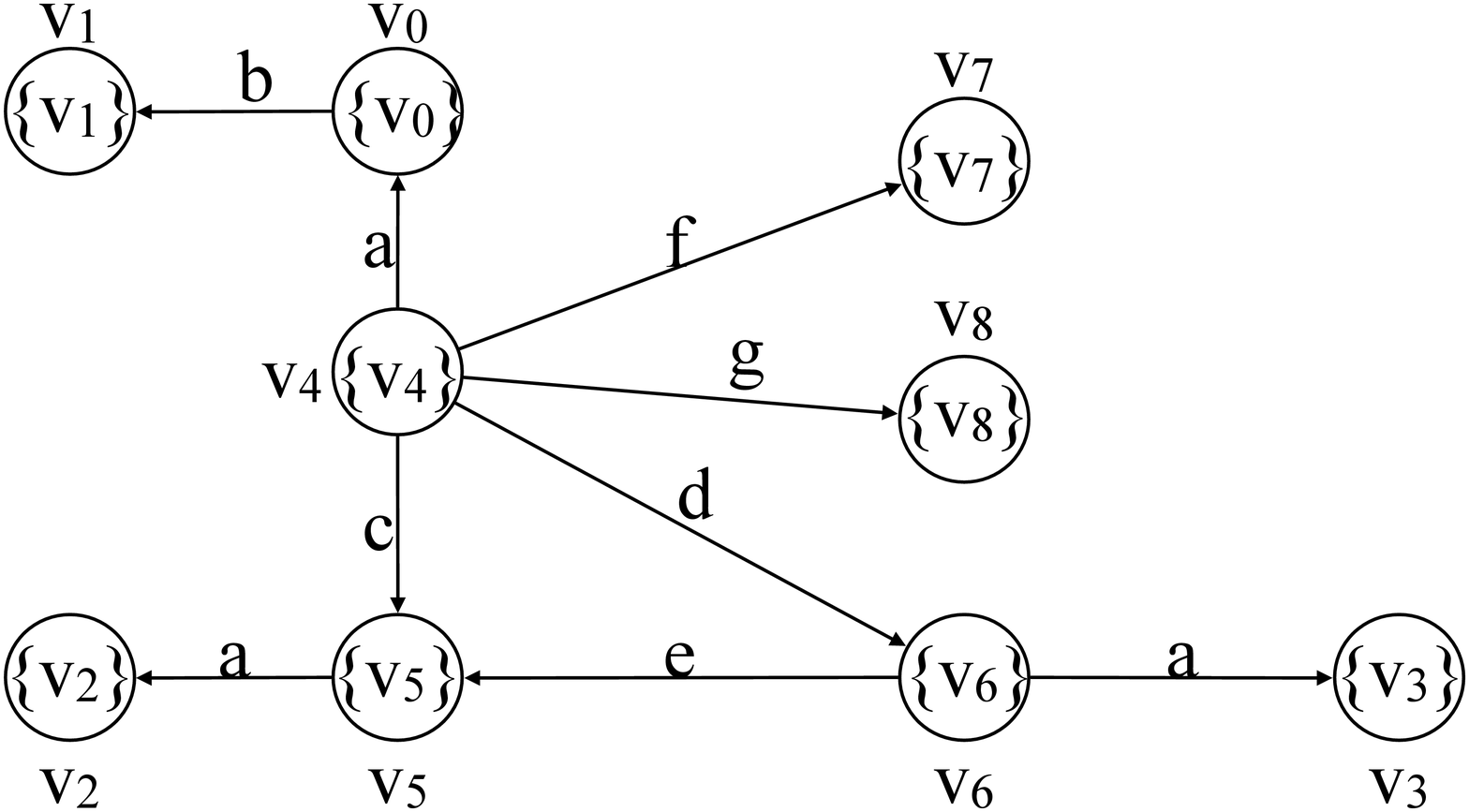}
  }
  \vspace{-1mm}
  \caption{Direct transformation of RDF graph (Vertex label function $L(v) = \{v\}$).}
  \vspace{-2mm}
  \label{fig:direct-transformation}
\end{figure}

A query graph is obtained from a SPARQL query. A query vertex may hold
the vertex label which corresponds to the subject or object
specified in the SPARQL query. If the query vertex corresponds to a
variable, the vertex label is left blank. For example, the
SPARQL query of Figure~\ref{fig:direct-trans-sparql-q2} is transformed
into the query graph of Figure~\ref{fig:direct-trans-query-graph-q2}.
Here the query vertex $u_0$, which corresponds to \verb|Student|,
holds the vertex label $\{v_1\}$; To the contrary, the query vertex
$u_3$, which corresponds to the variable $X$, has blank (\verb|_|) as
the vertex label. Similarly, a query edge may hold the edge
label which corresponds to the predicate. For example, the edge label
of $(u_3, u_4)$ is $c$ as the edge corresponds to the
\verb|undergradDegreeFrom| predicate.

\begin{figure}[h!]
  \vspace{-2mm}
  \centering

  \begin{myverbbox}{\vqtwo}
SELECT ?X, ?Y, ?Z WHERE
{?X rdf:type Student .
 ?Y rdf:type University .
 ?Z rdf:type Department .
 ?X undergradDegreeFrom ?Y .
 ?X memberOf ?Z .
 ?Z subOrganizationOf ?Y.}
    \end{myverbbox}
  \subfloat[SPARQL query.]
  {
    \label{fig:direct-trans-sparql-q2}
    \resizebox{0.5\linewidth}{!}{ \raisebox{5mm}{\vqtwo}}
  }
  \subfloat[query graph.]
  {
    \label{fig:direct-trans-query-graph-q2}
    \includegraphics[width=0.45\linewidth]{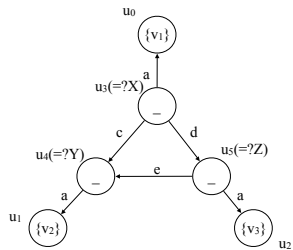}
  }
  \vspace{-1mm}
  \caption{Direct transformation of SPARQL query.}
  \vspace{-1mm}
  \label{fig:direct-transformation-query}
\end{figure}

Note that, when a variable is declared on a predicate in a SPARQL
query, a query edge has a blank edge label. An e-graph homomorphism
algorithm can answer such SPARQL queries since an e-graph homomorphism
has edge label mapping from query edges to their corresponding edge
labels.

Consequently, the direct transformation makes it possible to apply
conventional e-graph homomorphism algorithms for processing SPARQL
queries. In order to evaluate the performance of such
an approach, we applied \TurboISOp{} on \edit{LUBM8000, a
  billion-triple RDF dataset of Leihigh University Benchmark (LUBM)
  \cite{GPH:05aa}}, after applying direct transformation. We compared
the performance of \TurboISOp{} against two existing RDF engines:
RDF-3X \cite{NW:10aa}, and System-X\footnote{We anonymize the product
  name to avoid any conflict of interest.}.
Figure~\ref{fig:comparison-direct-trans} depicts the measured
execution time of these three systems in log scale. (See
Section~\ref{sec:experiment-setup} for the details of the experiment
setup)

\begin{figure}[h!]
  \vspace{-1mm}
  \centering
  \includegraphics[width=\linewidth]{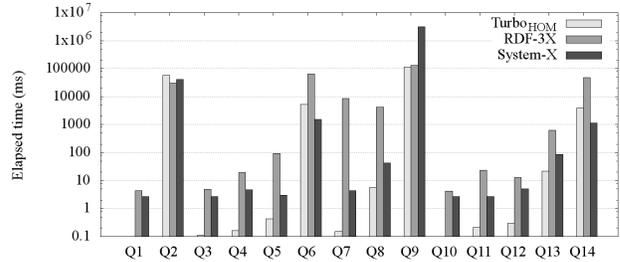}
  \vspace{-6mm}
  \caption{Comparison between original \TurboISOp{} with the direct
    transformation graph and other RDF engines.}
  \label{fig:comparison-direct-trans}
\end{figure}

Although there is no clear winner among them, the figure reveals that
\TurboISOp{} performs as good as the existing RDF engines. For
short-running queries (i.e Q1, Q3-Q5, Q7, Q8, Q10-Q13), \TurboISOp{}
shows faster elapsed time. As those queries specify a data vertex ID,
\TurboISOp{} only needs a small amount of graph exploration from one
candidate region with an optimal matching order, while RDF-3X and
System-X require expensive join operations. For long-running queries
(i.e., Q2, Q6, Q9, and Q14), \TurboISOp{} is slower than some of its
competitors. The performance of \TurboISOp{} largely relies on 1)
graph exploration by $ExploreCan\-didateRegion$ and 2) subgraph
enumeration by $SubgraphSearch$. Moreover, when a query graph has
non-tree edges, $IsJoinable$ constitutes a large portion of
$SubgraphSearch$. The profiling results of long running queries
confirmed that 1) $ExploreCandidate\-Region$ and $SubgraphSearch$ are
the dominating factors and 2) for queries which have non-tree edges
(Q2 and Q9), $IsJoinable$ is the dominating factor of
$Subgraph\-Search$.  Specifically, \TurboISOp{} spent the most time on
$Explore\-Candidate\-Region$ (e.g. 46\% for Q2, 70\% for Q6, 72\% for
Q9, and 69\% for Q14) and $Subgraph\-Search$ (e.g. 54\% for Q2, 30\% for
Q6, 28\% for Q9, and 31\% for Q14). Moreover, for queries which have
non-tree edges, the most of $Sub\-graphSearch$ time was spent on
$IsJoinable$ (e.g. 81.4\% for Q2 and 77.6\% for Q9).
\edit{
  In order to speed up $Explore\-Candidate\-Region$, we propose a novel
  transformation (Section~\ref{sec:type-awre-transf}). Tailored optimization
  techniques are proposed for improving performance for both functions (Section~\ref{sec:optimization}).}

\section{Turbo\textsf{HOM}++}
\label{sec:turbohom++}

In this section, we propose an improved e-graph homomorphism
  algorithm, \TurboISOP{}. Introduced first is the \textit{type-aware
    transformation}, which can result in faster pattern matching than
  direct transformation
  (Section~\ref{sec:type-awre-transf}). \TurboISOP{} processes the labeled
  graph transformed by the type-aware transformation
  (Section~\ref{sec:data-structure}). Furthermore, for efficient RDF
  query processing, four optimizations are applied to \TurboISOP{}
  (Section~\ref{sec:optimization}).

\subsection{Type-aware Transformation}
\label{sec:type-awre-transf}

To enable the type-aware transformation, we devise the
\textit{two-attribute vertex model} which makes use of the type
information specified by the \verb|rdf:type| predicate. Specifically,
this model assumes that each vertex is associated with a set of labels
(the label attribute) in addition to its ID (the ID attribute). The
label attribute is obtained by following the \verb|rdf:type| predicate
-- if a subject has one or more \verb|rdf:type| predicates, its types
can be obtained by following the \verb|rdf:type| (as well as
\verb|rdf:subClassOf| predicates transitively). For example,
\verb|student1| in Figure~\ref{fig:graph} has the label attribute,
$\{\verb|GradStudent|,$ $\verb|Student|\}$.

The above two-attribute vertex model naturally leads to our new RDF
graph transformation, \textit{the type-aware transformation}. Here,
subjects and objects are transformed to two-attribute vertices by
utilizing \verb|rdf:type| predicates as described above. Then, the ID
attribute corresponds to the vertex ID, and the label attribute
corresponds to the vertex label.
Figure~\ref{fig:type-aware-transformation} shows an example of the
mapping tables and the data graph, which is the result of type-aware
transformation applied to Figure~\ref{fig:graph}. \edit{Now, we
  formally define the type-aware transformation as follows.}

\begin{definition}
  \label{def:type-aware}
  \edit{The type-aware transformation
    $(F_V, F_{ID}, F_E, F_{VL},$ $F_{EL})$ converts a set of triples
    $T(S, P, O)$ to a type-aware transformed graph $G(V,E,ID,L)$.  Let
    us divide $T$ into three disjoint subsets whose union is $T$ ---
    $T'(S',P',O')$,
    $T'_{\mathsf{t}}(S'_{\mathsf{t}}, P'_{\mathsf{t}},
    O'_{\mathsf{t}}) = \{(s,\text{rdf:type},o) \in T\}$,
    and
    $T'_{\mathsf{sc}}(S'_{\mathsf{sc}}, P'_{\mathsf{sc}},
    O'_{\mathsf{sc}}) = \{(s,\text{rdf:subClassOf},$ $o) \in T \}$.}

  \begin{enumerate}[1.,leftmargin=0.4cm]
    \setlength{\itemsep}{-3pt}
  \item A vertex mapping $F_V: S' \cup O' \cup S'_\mathsf{t} \to V$,
    which is bijective, maps a subject in $S' \cup S'_\mathsf{t}$ or
    an object in $O'$ to a vertex.
  \item A vertex ID mapping
    $F_{ID}: S' \cup O' \cup S'_\mathsf{t} \to N \cup \{\_\}$, which
    is bijective, maps a subject in $S' \cup S'_\mathsf{t}$ or an
    object in $O'$ to a vertex ID or blank. Here, $F_{ID}(x) = \_$ if x
    is a variable.
  \item An edge mapping $F_E: T' \to E$, which is bijective, maps a
    triple of $T'$ into an edge, $F_E(s,p,o) = (F_V(s), F_V(o))$.
  \item A vertex label mapping
    $F_{VL} : O'_\mathsf{t} \cup O'_{\mathsf{sc}} \to VL \cup \{\_\}$, which is
    bijective, maps an object of $O'_\mathsf{t} \cup O'_{\mathsf{sc}}$
    into a vertex label. Here, $F_{VL}(x) = \_$ if x is a variable.
  \item An edge label mapping $F_{EL} : P' \to EL \cup \{\_\}$, which is
    bijective, maps a predicate of $P'$ into an edge label. Here, $F_{EL}(x) = \_$ if x is a variable.
  \item A vertex ID mapping function $ID: V \to N$ maps a
    vertex to a vertex ID where
    $ID(v) = F_{ID} \circ F_V^{-1}(v)$.
  \item A labeling function $L$ 1) maps a vertex to a set of vertex labels such that $v \in V$,
    $L(v) = \{F_{VL}(o) | \text{ there is a path from } F_V^{-1}(v)$
	$\text{to }$ $o \text{ using triples in } T'_{\mathsf{t}} \cup T'_{\mathsf{sc}}\}$ and 2) maps an edge e to an edge label such that $e \in E$, $L(e) = F_{L_E}(Pred(F_E^{-1}(e)))$ where
	$Pred(s,p,o) = p$.
  \end{enumerate}
\end{definition}


\setlength{\tabcolsep}{0.7mm}
\begin{figure}[h!]
  \centering
  \subfloat[vertex ID mapping table.]
  {
    \label{fig:type-aware-vid-map}
    \scriptsize
    \begin{tabular}[t]{|c|c|}
      \hline
      Subject/Object & Vertex ID \\
      \hline
       student1 & $0$ \\
       univ1 & $1$ \\
       dept1.univ1  & $2$ \\
       `012-345-678' & $3$ \\
       `john@dept1.univ1.edu' & $4$ \\
      \hline
    \end{tabular}
  }
  \hspace*{0.15cm}
  \subfloat[vertex label mapping table.]
  {
    \label{fig:type-aware-vlabel-map}
    \scriptsize
    \begin{tabular}[t]{|c|c|}
      \hline
      Type & Vertex Label \\
      \hline
       GraduateStudent & A \\
       Student & B \\
       University & C \\
       Department & D \\
      \hline
    \end{tabular}
  }
  \\
  \subfloat[edge label mapping table.]
  {
    \label{fig:type-aware-elabel-map}
    \scriptsize
    \begin{tabular}[b]{|c|c|}
      \hline
      Predicate & Edge Label \\
      \hline
       undergradDegreeFrom & a \\
       memberOf & b \\
       subOrganizationOf & c \\
       telephone & d \\
       emailAddress & e \\
      \hline
    \end{tabular}
  }
  \subfloat[data graph.]
  {
    \label{fig:type-aware-trans-graph}
    \includegraphics[width=0.4\linewidth]{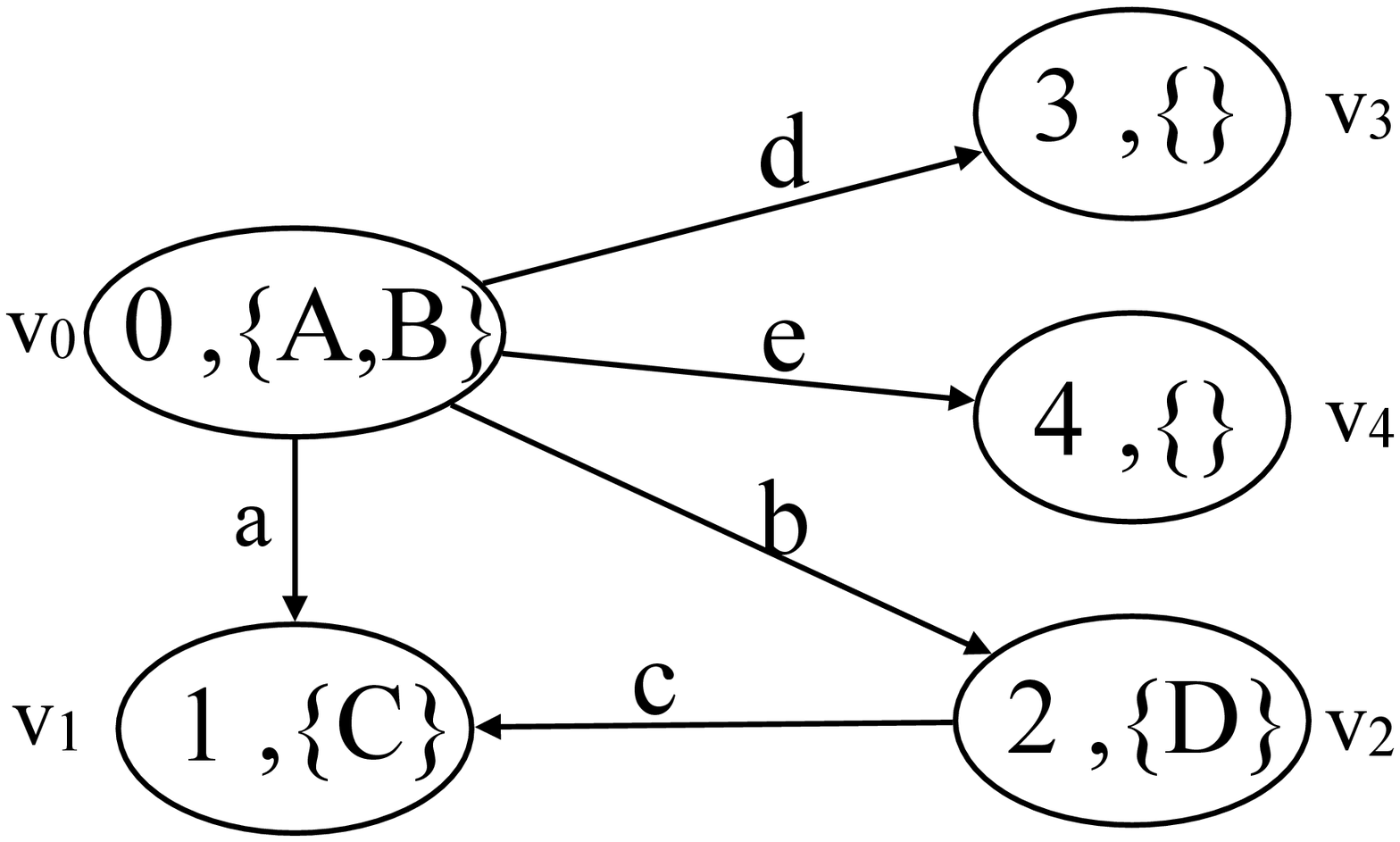}
  }
  \caption{Type-aware transformation of an RDF graph.}
  \label{fig:type-aware-transformation}
\end{figure}

After finding a type-aware transformation
$(F'_V, F'_{ID}, F'_E, F'_{VL},$ $k F'_{EL})$ for a data graph
$g(V',E',L',ID')$, we can also convert a SPARQL query into a
type-aware transformed query graph $q(V,E,$ $L,ID)$ by using another
type-aware transformation $(F_V, F_{ID}, F_E,$ $F_{VL}, F_{EL})$ such
that $F_{ID} = F'_{ID}$, $F_{VL} = F'_{VL}$, and $F_{EL} =
F'_{EL}$.
For example, Figure~\ref{fig:type-ware-trans-query-q2} is the query
graph type-aware transformed from the SPARQL query in
Figure~\ref{fig:direct-trans-sparql-q2}. Note that a query vertex may
have multiple vertex labels like a data vertex.

\begin{figure}[h!]
  \centering
  \includegraphics[width=0.35\linewidth]{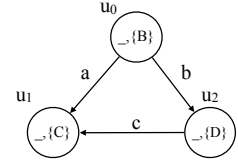}
  \caption{Type-aware transformation of SPARQL query
    of Figure~\ref{fig:direct-trans-sparql-q2}.}
  \vspace{-3mm}
  \label{fig:type-ware-trans-query-q2}
\end{figure}

\edit{Now, we explain how the generic e-graph homomorphism
  algorithm works for type-aware transformed query/data graphs.
  When appending a candidate data vertex to the current partial
  solution, we additionally check the following condition for the ID
  attribute of the two-attribute vertex model.

  \[ \forall u \in \{u | ID(u) \neq \_ \text{ for } u \in V \}, ID(u)
  = ID'(M_v(u)).\]}

The virtue of the type-aware transformation is that it can improve the
efficiency of RDF query processing. Since the type-aware
transformation eliminates certain vertices and edges by embedding {\it
  type} information into the vertex label, the resulting data/query
graphs have smaller size and simpler topology than those transformed
by the direct transformation.

As an example, let us consider the SPARQL query in
Figure~\ref{fig:direct-trans-sparql-q2}. After direct transformation,
it becomes the query graph in
Figure~\ref{fig:direct-trans-query-graph-q2} that has a relatively
complex topology consisting of six vertices and six edges. On the
other hand, the type-aware transformation produces the query graph in
Figure~\ref{fig:type-ware-trans-query-q2} that has a simple triangle
topology. This reduced number of vertices and edges has a positive
effect on efficiency because it results in less graph exploration.

In general, the effect of the type-aware transformation can be
described in terms of the number of data vertices in all candidate
regions. \editt{Consider a SPARQL query which consists of a set of
  triples $T$, its direct transformed query graph $q(V,E,L)$, and its
  type-aware transformed query graph $q'(V',E',ID',L')$. Let
  $O_{type} = \{o | (s, \mathsf{rdf:type}, o) \in T \text{ or } (s,
  \mathsf{rdf:subClassOf}, o) \in T\}$.
  In the direct transformation, $o \in O_{type}$ is transformed to a
  query vertex. Let $V_{type}$ a set of direct transformed query
  vertices from $O_{type}$. However, in the type-aware transformation,
  $o \in O_{type}$ is not transformed to a query vertex, which
  satisfies $|V'| = |V| - |V_{type}|$. Therefore, the type-aware
  transformation leads to less graph exploration in
  $ExploreCandidateRegion$ and $SubgraphSearch$. Formally, using the
  type-aware transformation, the number of data vertices in all
  candidate regions is reduced by
  \[ \sum_{v_s} \sum_{u \in V_{type}} |CR_{v_s}(u)|\]
  where $v_s$ represents the starting data vertex for each candidate
  region, and $CR_{v_s}(u)$ represents a set of data vertices in a
  candidate region $CR(v_s)$ that correspond to $u$. }
\subsection{Implementation}
\label{sec:data-structure}

\TurboISOP{} maintains two in-memory data structures -- the inverse
vertex label list and the adjacency list.
Figure~\ref{fig:inverse-vertex-list} shows the inverse vertex label
list of Figure~\ref{fig:type-aware-trans-graph}. The `end offsets'
records the exclusive end offset of the `vertex IDs' for each vertex
label. Figure~\ref{fig:adjacency-list} shows the adjacency list of
Figure~\ref{fig:type-aware-trans-graph} for the outgoing edges. The
adjacency list stores the adjacent vertices for each data vertex in
the same way as the inverse vertex label list. One difference is that
the adjacency list has an additional array (`end offsets') to group
the adjacent vertices of a data vertex for each neighbor type. Here,
the neighbor type refers to the pair of the edge label and the vertex
label. For example, $v_0$ in Figure~\ref{fig:type-aware-trans-graph},
has four different neighbor types -- $(a,C), (b,D), (d,\_)$ and
$(e,\_)$. Those four neighbor types are stored in `end offsets,' and
each entry points to the exclusive end offset of the `adjacent vertex
ID'. \TurboISOP{} maintains another adjacency list for the incoming
edges.

\edit{We assume that graphs in our system are periodically updated from an underlying RDF source. For efficient graph update, a
  transactional graph store is definitely required. We leave this
  exploration to future work since it is beyond the scope of the
  paper.}

\edit{Note also that \TurboISOP{} can also handle SPARQL queries
  under the simple entailment regime correctly. In order to deal with
  the simple entailment regime in the type-aware transformed graph,
  \TurboISOP{} distinguishes
  $L_{simple}(v) = \{F_{L_V}(o) | \text{there is an edge}$ $ \text{from }
  F_V^{-1}(v) \text{ to } o \text{ using triples in } T'\}$
  from $L(v)$. \TurboISOP{} can process a SPARQL query under the
  simple entailment regime using $L_{simple}(v)$ instead of
  $L(v)$. \\}

\begin{figure}[h!]
  \vspace{-3mm}
  \centering
  \subfloat[inverse label vertex list.]
  {
    \label{fig:inverse-vertex-list}
    \includegraphics[width=0.43\linewidth]{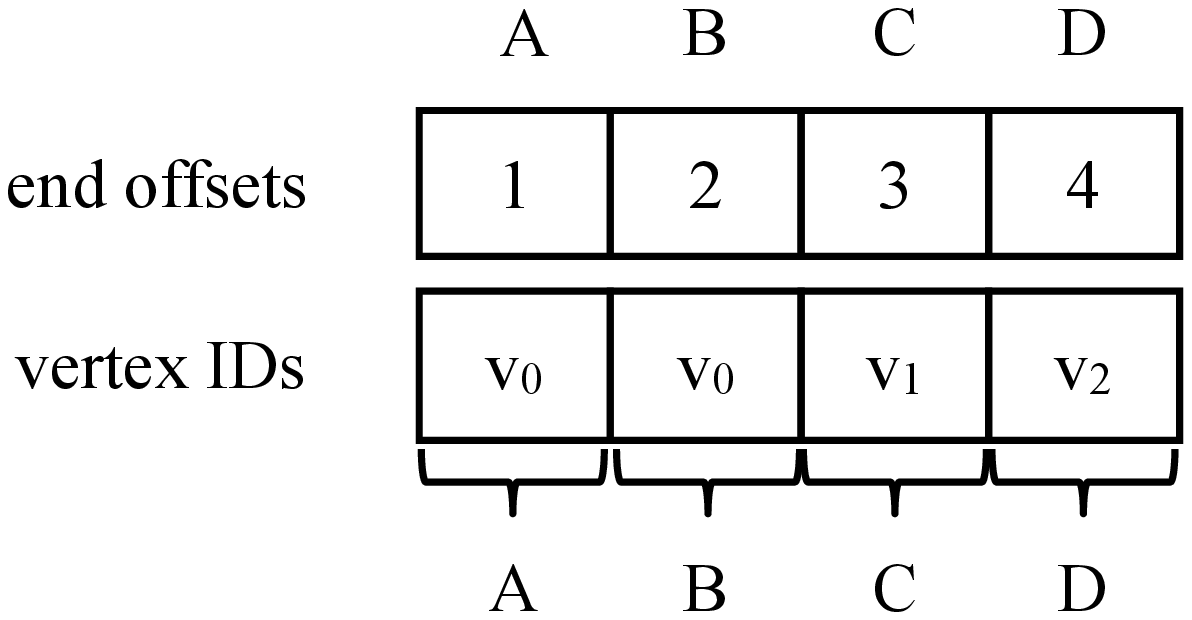}
  }
  \\
  \subfloat[adjacency list.]
  {
    \label{fig:adjacency-list}
    \includegraphics[width=0.9\linewidth]{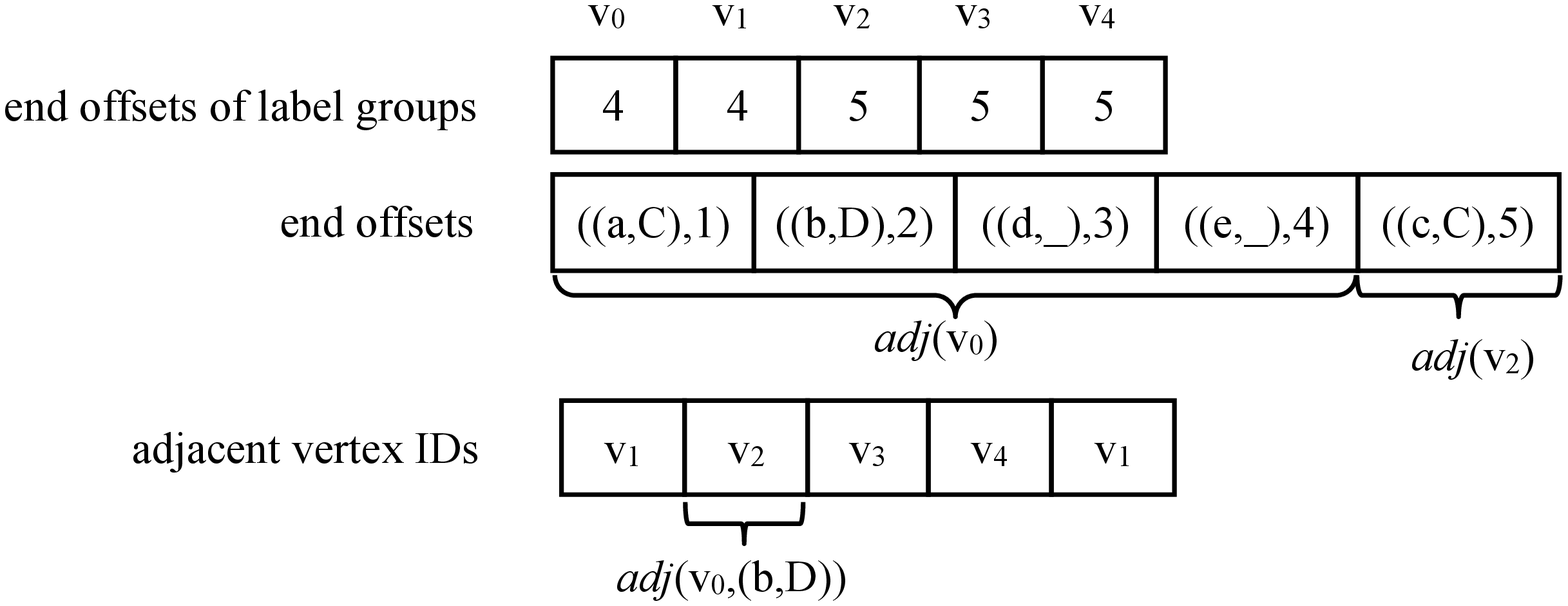}
  }
  \caption{In-memory data structures for type-aware transformed
    data graph of Figure~\ref{fig:type-aware-trans-graph} ($adj(v):$
    adjacent vertices of $v$, $adj(v,(el,vl)):$ adjacent vertices $v$,
    which have vertex label $vl$ and are connected with edge label
    $el$).}
  \label{fig:datastructure}
\end{figure}

As the overall behavior of \TurboISOP{} is similar to
  \TurboISOp{}, here, we describe how \TurboISOP{} uses the data
  structures in $Choose\-Start\-Query\-Vertex$ (line 6 of
  Algorithm~\ref{alg:turboiso}), $Explore\-Candidate\-Region$ (line 9
  of Algorithm~\ref{alg:turboiso}), and $IsJoinable$ (line 7 of
  Algorithm~\ref{alg:subgraphsearch}).

  \textbf{ChooseStartQueryVertex. } When computing $rank(u)$ for a \\
  query vertex $u$, the inverse vertex list is used to get
  $freq(g, L(u))$ (= $|\bigcap_{l \in L(u)} V(g)_{l}|$) where $V(g)_l$
  is the set of vertices having vertex label $l$. When $|L(u)| = 1$,
  Getting the start and end offset of a specific vertex
  label is enough. When $|L(u)| > 1$, for each $l \in |L(u)|$, all data vertices
  having $l$, $V(g)_{l}$, are retrieved from the inverse vertex list,
  and $freq(g, L(u))$ is obtained by intersecting all
  $V(g)_{l}$. Additionally, when a data vertex ID $v$ is specified in
  $u$, $freq(g, L(u)) = 1$ if $v \in V(g)_l$ for each $l \in
  L(u)$. Otherwise, $freq(g, L(u)) = 0$.

One last case is when a SPARQL query has a query vertex which has no
label or \text{ID} at all. In order to handle such queries, we
maintain an index called the predicate index where a key is a
predicate, and a value is a pair of a list of subject IDs and a list
of object IDs. This index is used to compute $freq(g, L(u))$.

\textbf{ExploreCandidateRegion. }After a query tree is
  generated, candidate regions are collected by exploring the data
  graph in an inductive way. In the base case, all data vertices that
  correspond to the start query vertex are gathered in the same way of
  computing $freq(g, L(u))$. In the inductive case, once the starting
  data vertices are identified, the candidate region exploration
  continues by exploiting the adjacency information stored in the
  adjacency list. If one vertex label and one edge label are specified
  in the query graph, we can get the adjacent data vertices directly
  from the adjacency list. If multiple vertex labels and one edge
  label are specified, we collect the adjacent data vertices for each
  vertex label using the adjacency list, and intersect them. In a case
  where the vertex label or edge label is blank, \TurboISOP{} finds the
  correct adjacent data vertices by 1) collecting all adjacent
  vertices which match available information (either vertex label or
  edge label) and 2) unioning them. Additionally, if the current query
  vertex has the data vertex ID attribute, we check whether the
  specified data vertex is included in the data vertices collected
  from the adjacency list.

\textbf{IsJoinable. }The $IsJoinable$ test is equivalent to
  the inductive case of $ExploreCandidateRegion$ when a data vertex ID
  (previously matched data vertex) is specified.

\subsection{Optimization}
\label{sec:optimization}

In this subsection, we introduce optimizations that we apply to improve
the efficiency of \TurboISOP{}.
Even though these optimizations do not change
\TurboISOP{} severely, they could improve the query processing
efficiency quite significantly.


\textbf{Use intersection on $IsJoinable$ test \textsf{(+INT)}. }
\sloppy We optimize the $IsJoinable$ test in $SubgraphSearch$.
$SubgraphSearch$ calls the $IsJoinable$ test by multiple membership
operations. However, the optimization allows a bulk of $IsJoinable$
tests with one $k$-way intersection operation where $k$ is the number
of edges between the current query vertex, $u$ in line 1 of
Algorithm~\ref{alg:subgraphsearch}, and the previously matched query
vertices connected by non-tree edges.

  $SubgraphSearch$ checks the existence of the edges between the
  current candidate data vertex and the already bounded data vertices
  by calling $IsJoinable$ (line 7 of
  Algorithm~\ref{alg:subgraphsearch}) when the corresponding query
  graph has non-tree edges. Let us consider the query graph
  (Figure~\ref{fig:type-ware-trans-query-q2}), the query tree
  (Figure~\ref{fig:nec-q2}) and a data graph
  (Figure~\ref{fig:opt-intersection}). Suppose that, for a given
  matching order $u_1 \to u_2 \to u_0$, the vertex $v_1$ is bound to
  $u_1$, and the vertex $v_2$ is bound to $u_2$. Then, the next step
  is to bind a data vertex to $u_0$. Because there is a
  non-tree edge between $u_0$ and $u_2$, to bind a data vertex
  of ID $v_i (i =0,3,4,\cdots,1001$) to $u_0$, we need to check
  whether there exists an edge $v_i \to v_2$.

\begin{figure}[h!]
  \centering
  \includegraphics[width=0.3\linewidth]{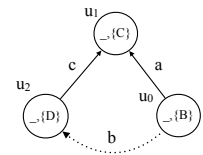}
  \caption{A query tree of the query graph of
    Figure~\ref{fig:type-ware-trans-query-q2}.}
  \label{fig:nec-q2}
\end{figure}

\begin{figure}[h!]
  \centering
  \label{fig:data-graph}
  \includegraphics[width=0.65\linewidth]{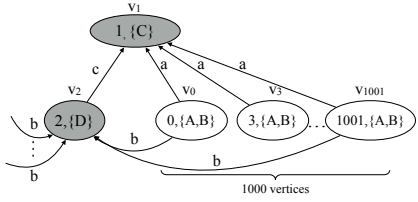}
  \vspace{-1mm}
  \caption{An example data graph for illustrating \textsf{+INT}.}
  \vspace{-2mm}
  \label{fig:opt-intersection}
\end{figure}

$IsJoinable$ checks for the existence of the edge between the current
data vertex and already matched data vertices by repetitively calling
$IsJoinable$. Let us consider the above example. For each $v_i (i =
0,3,4,\cdots,1001)$, $IsJoinable$ tests whether the edge $v_i \to v_2$
exists. If $v_2$ is a member of $v_i$'s outgoing adjacency list, the
test succeeds, and the graph matching continues.

Instead, our modified $IsJoinable$ tests all the edge occurrences
between the current candidate vertices ($C_{R}$ in line 3 of
Algorithm~\ref{alg:subgraphsearch}) and the adjacency lists of the
already matched data vertices by one $k$-way intersection operation.
Let us consider the above example again. The modified $IsJoinable$
finds the edge between $v_2$ and the candidate data vertices
$v_0,v_3,\cdots,v_{1001}$ at once. For this, it is enough to perform
one intersection operation between the $v_2$'s incoming adjacency
vertices and the candidate data vertices. Since the modified
$IsJoinable$ takes $C_R$ as a parameter, the lines 3 and 7 of
Algorithm~\ref{alg:subgraphsearch} are merged into one statement.

Note that this optimization can improve the performance significantly.
In the above example, since only $v_0$ and $v_{1001}$ pass the test,
we can avoid calling the original $IsJoinable$ $998$ times. Formally
speaking, let us denote 1) the candidate data vertex set for the
current query vertex $u$ as $C_{R}$,
\editt{2) the previously matched query
vertex set, which is connected to the current query vertex by non-tree
query edges, as $\{u'_i\}_{i=1}^{k}$ and 3) the adjacent vertex set
of $v'_i(= M_v(u'_i))$ where $u'_i$ is connected to $u$ with the
vertex label $vl_i$ and the edge label $el_i$, as $adj(v'_i, vl_i,
el_i)$. Suppose that $C_{R}$ and $adj(v'_i, vl_i, el_i)$ are stored
in ordered arrays. Then, the complexity of the original $IsJoinable$
test is
\[ C_{original} = O( |C_{R}| \cdot \sum_{i=1}^k \log |adj(v'_i, vl_i,
el_i)|) \],
since $IsJoinable$ is called for each $v \in C_R$, and $O(\log |adj$
$(v'_i, vl_i, el_i)|)$ time is required to conduct a binary search for
$|adj(v'_i, vl_i, el_i)|$ elements.  On the contrary, the complexity
of the modified $IsJoinable$ test is
  \[ \min( O(|C_{R}| + \sum_{i=1}^k |adj(v'_i, vl_i, el_i)|),
  C_{original})\] since the modified $IsJoinable$ can choose the
  $k$-way intersections strategy between scanning $(k+1)$ sorted lists
  and performing binary searches.} 
\textbf{Disable NLF Filter \textsf{(-NLF)}. }\edit{The second
  optimization is to disable the NLF filter in
  $ExploreCandidateRegion$. The NLF filter may be effective when the
  neighbor type are very irregular. However, in practice, most RDF
  datasets are structured~\cite{NM:11aa,GN:14aa}. For example, in our
  sample RDF dataset (Figure~\ref{fig:graph}), in most case, a vertex
  corresponding to a graduate student has} \verb|telephone|,
\verb|emailAddress|, \verb|memberOf|, and
\verb|undergraduateDegreeFrom| \edit{predicates. Accordingly, the NLF
  filter is not helpful for such structured RDF datasets.}



\textbf{Disable Degree Filter \textsf{(-DEG)}.} \edit{The third
  optimization is to disable the degree filter in
  $ExploreCandidateRegion$. Similar to the NLF filter, the degree
  filter is effective when the degree is very irregular while RDF
  datasets typically are not.}



\textbf{Reuse Matching Order (\textsf{+REUSE}). }
The last optimization is to reuse the matching order of the first
candidate region for all the other candidate regions. That is,
$DetermineMatchingOrder$ (line 6 of
Algorithm~\ref{alg:turboiso}) is called only once throughout the
\TurboISO{} execution, and the same matching order is used throughout
the query processing. \TurboISOP{} uses a different matching order for
each candidate region, because each candidate region could have a very
different number of candidate vertices for a given query path in the
e-graph homomorphism problems. However, \edit{typical RDF datasets are regular
at the schema level, i.e. well structured in practice,} and generating the
matching order for each candidate region is ineffective, especially
when the size of each candidate region is small. \edit{We also performed
experiments with more heterogeneous datasets, including Yet
Another Great Ontology (YAGO) \cite{SKW:08aa}, and Billion Triples Challenge 2012 (BTC2012) \cite{H:12aa}.
This optimization technique still shows good matching performance as we will see in our extensive experiments in Section~\ref{sec:experiment}, since these heterogeneous datasets do not show extreme irregularity at the schema level.}


\iftr
Let us take the example of the expanded RDF data from
Figure~\ref{fig:graph} and the query graph of
Figure~\ref{fig:type-ware-trans-query-q2}. Suppose that the query tree
is Figure~\ref{fig:nec-q2}. In that case, the starting data vertices
of the candidate regions are the data vertices with the
\verb|University| vertex label. To make a candidate region for a
starting data vertex, \verb|univ1|, we must find (1) all departments
in \verb|univ1| and (2) all the graduate students who got their
undergraduate degrees from \verb|univ1|. Because the selectivity of
(2) is higher than that of (1), the matching order, $u_1 \to u_2 \to
u_0$ is chosen. For the other universities (starting vertices), it is
rare that the selectivity of (2) is higher than that of (1). For such
a case, it is more efficient to reuse the first matching order.
\fi

\iftr
\section{Further Improvement Of Turbo$_{HOM}$++}
\label{sec:extension}

In this section, we briefly describe how \TurboISOP{} can handle the
general SPARQL keywords (Section~\ref{sec:pattern-modifier}), and how
it can be parallelized under NUMA architecture
(Section~\ref{sec:parallel-processing}).

\subsection{Supporting General SPARQL Keywords}
\label{sec:pattern-modifier}

Along with the basic graph pattern matching, we
  briefly describe how an e-graph homomorphism algorithm can handle
  the general SPARQL keywords -- OPTIONAL, FILTER, and UNION. 
  Thus, \TurboISOP{} can support the explore use case queries of the Berlin
  SPARQL benchmark \cite{BS:09aa} using OPTIONAL, FILTER, and
  UNION.

\textbf{OPTIONAL.}\edit{ To support queries that ask information which is not
  necessarily required, the OPTIONAL keyword is
  used. Figure~\ref{fig:optional} is an example of such a query which
  finds the price of} \verb|<product1>| \edit{and its rating and its
  homepages if possible. To handle OPTIONAL in \TurboISOP{}, we propose a
  simple yet effective technique as follows. First, \TurboISOP{}
  selects a start query vertex which is not specified in an OPTIONAL
	  clause. Then,
  \TurboISOP{} makes a candidate solution using the
  \emph{nullify-and-keep-searching} strategy. In
  $ExploreCandidateRegion$, if the current query vertex is in an OPTIONAL clause,
  and no data vertex is matched, \TurboISOP{} nullifies the current
  query vertex mapping in a candidate region. In $SubgraphSearch$,
  even though the mapped data vertex is nullified, if the
  corresponding query vertex is in an OPTIONAL clause, it invokes a recursive
  call. After a candidate solution is constructed using
  nullify-and-keep-searching, \TurboISOP{} qualifies it using the
  \emph{qualify-and-exclude-duplicate} strategy. The OPTIONAL semantics
  enforces that all vertices in an OPTIONAL clause must be mapped to data vertices,
  otherwise, the mappings of all query vertices in an OPTIONAL clause are
  nullified. Also, when all vertices in an OPTIONAL clause are nullified, the
  nullified final mapping should be generated only once. \TurboISOP{}
  excludes the duplicates by comparing the current final mapping with
  the previous valid mapping. For example, suppose two successive
  solutions of the example query are}
$\{$\verb|(price,$100)|, \verb|(rating, 5)|,
\verb|(homepage,null)|$\}$,
\textit{and} $\{$\verb|(price,$100)|,
\verb|(rating, 1)|, \verb|(homepage,null)|$\}$. \edit{The preceding
  final solution is qualified as}
$\{$\verb|(price,$100)|, \verb|(rating, null)|, \verb|(homepage,|
\verb| null)|$\}$,
\edit{but the latter solution is dropped because it is the same as the
  preceding final solution. The qualify-and-exclude-duplicate
  strategy is recursively applied to handle the nested OPTIONAL
  clauses.}
\begin{figure}[h!]
  \centering
  \begin{myverbbox}{\voptional}
SELECT ?price ?rating ?homepage WHERE
{ <product1> rdf:type <Product>. <product1> price ?price.
 OPTIONAL  {<product1> rating ?rating.
            <product1> homepage ?homepage.} }
  \end{myverbbox}
  \resizebox{1\linewidth}{!}{\voptional}
  \vspace{-0.7cm}
  \caption{A SPARQL query which has \edit{an} OPTIONAL keyword.}
  \label{fig:optional}
\end{figure}

\textbf{FILTER.} To restrict solutions that do not qualify conditions, the FILTER
keyword is used. Figure~\ref{fig:filter} is an example of such a query
which finds all products which have a higher rating than
\verb|<product1>|. To handle FILTER expressions, inexpensive filters
such as selection conditions are applied whenever we access the
corresponding vertices, while expensive filters such as join
conditions and regular expressions are applied after we find a
solution without these expensive filters.


\begin{figure}[h!]
  \centering
  \begin{myverbbox}{\vfilter}
SELECT ?product WHERE
{ <product1> rdf:type <Product>. <product1> rating ?r1.
  ?product rdf:type <Product>. ?product rating ?r2.
  FILTER(?r2 > ?r1) }
  \end{myverbbox}
  \resizebox{1\linewidth}{!}{\vfilter}
  \vspace{-0.7cm}
  \caption{A SPARQL query which has a FILTER keyword.}
  \label{fig:filter}
\end{figure}

\textbf{UNION. }In SPARQL, to support the alternative pattern
matching, the UNION keyword is used. Figure~\ref{fig:union} is an
example of such a query which finds products having either
\verb|<feature1>| or \verb|<feature2>|. To handle the UNION keyword,
the SPARQL query is split into sub-queries, and an e-graph
homomorphism algorithm solves each sub-query. Then, the final
solutions are the union of the sub-queries' solutions, as the semantic
of the UNION keyword does not remove duplicated items.

\begin{figure}[h!]
  \centering
  \begin{myverbbox}{\vunion}
SELECT ?product WHERE
{ {?product rdf:type <Product>. ?P hasFeature <feature1>.}
  UNION
  {?product rdf:type <Product>. ?P hasFeature <feature2>.} }
  \end{myverbbox}
  \resizebox{1\linewidth}{!}{\vunion}
  \vspace{-0.7cm}
  \caption{A SPARQL query which has a UNION keyword.}
  \label{fig:union}
\end{figure}
\fi
\iftr
\subsection{Parallel Processing}
\label{sec:parallel-processing}

After generating the query tree (line of
Algorithm~\ref{alg:turboiso}), each starting data vertex can be
processed independently -- including candidate region exploration,
matching order determination and subgraph search (lines 9 -- 15).
Therefore distributing a subset of the starting data vertices to each
thread is enough to parallelize \TurboISOP{}.

\textbf{Distributing Starting Data Vertices. }\edit{However, distributing the starting data vertices in a
  pre-determined way may lead to workload imbalance on threads. Although RDF datasets are
  regular at the schema-level, the cardinalities of one (many)-to-many
  relationships at the instance-level can significantly vary in candidate regions. This property
  even holds for joins in relational databases. For example, in
  Figure~\ref{fig:type-ware-trans-query-q2}, the query involves three
  types, University, Graduate students, and Departments at the schema level
  while universities can have significantly different numbers of
  graduated students and departments, which leads to different workload for each
  university vertex. To have as even a workload for each thread as
  much as possible, we assign a small chunk of the starting data
  vertices to threads dynamically.}

\textbf{NUMA-aware Parallelization. } \edit{The modern high-end
  workstations adopt the NUMA architecture to maximize the parallelism
  by using the multi-socket systems \cite{LPM:13aa,
    LBK:14aa}. However, to fully utilize parallelism NUMA provides, a
  parallel method should avoid remote memory access which retrieves
  data stored in a remote socket. To avoid that, first, each page of a
  data graph is allocated in sockets' local memory in a round-robin
  way. With this, each thread can expect uniform access latency for a
  data graph. Second, a thread is enforced to stick to a specific
  socket, and thread specific data structures are allocated in the
  same socket where the thread runs. By doing so, a thread accesses
  its own data structures without remote memory access.}




\fi

\section{Related Work}
\label{sec:related-work}


With the increasing popularity of RDF, the demand for SPARQL support in
relational databases is also growing. To meet such demand, most
open-source and commercial relational databases
support the RDF store and the RDF query processing.
RDF datasets are stored into relational tables with a set of indexes. After that, SPARQL queries are processed by translating
them into the equivalent join queries or by using special APIs.


To support RDF query processing, many specialized stores for RDF
data were proposed
\cite{NW:10aa,NW:10ab,YLW:13aa,ACZ:10aa,WW:06aa,BKH:02aa}. Similar to
RDBMS, RDF-3X \cite{NW:10ab,NW:10aa} treats RDF
triples as a big three-attribute table, but boosts the RDF query processing by
building exhaustive indexes
and maintaining statistics. RDF-3X
processes many SPARQL queries by using merge based join, which is efficient for disk-based and in-memory environments.
Different from RDF-3X, Jena \cite{WW:06aa} exploits multiple-property tables, while BitMat \cite{ACZ:10aa}
exploits 3-dimensional bit cube, so that it can also support 2D matrices of SO, PO, and PS.
H-RDF-3X \cite{HAR:11aa} is a distributed RDF processing engine where
RDF-3X is installed in each cluster node.


Several graph stores support RDF data in their native graph storages \cite{ZMC:11aa, ZYW:13aa}. gStore
\cite{ZMC:11aa} performs graph pattern matching using the
filter-and-refinement strategy. It first finds promising subgraphs using
the VS$^*$-tree index. After that, the exact subgraphs are
enumerated in the refinement step.
Trinity.RDF \cite{ZYW:13aa} is a sub-system of a distributed graph processing engine, Trinity \cite{SWL:13aa}.
The RDF triples are stored in Trinity's key-value store. When
processing RDF queries, Trinity.RDF implements special query processing methods for RDF data.


In 1976, \citet{U:76aa} published his seminal paper on the subgraph
isomorphism solution based on backtracking. After his work, many
subgraph isomorphism methods were proposed to improve the efficiency
by devising their own matching order selection algorithms and
filtering constraints
\cite{PFS:04aa,HS:08aa,SZL:08aa,ZLY:09aa,ZH:10aa,HLL:13aa}. Among
those improved methods, \TurboISO{} \cite{HLL:13aa} solves the
notorious matching order problem by generating the matching order for
each candidate region and by grouping the query vertices which have
the same neighbor information. The method shows the most efficient
performance among all representative methods.

Along with the backtracking based methods, the index-based subgraph
isomorphism methods were also proposed
\cite{CKN:07aa,YYH:04aa,YYH:05aa,ZHY:07aa,ZCY:08aa}. All of those
methods first prune out unpromising data graphs using low-cost filters
based on the graph indexes. After filtering, any subgraph isomorphism
methods can be applied to those unfiltered data graphs.  This
technique is only useful when there are many small data graphs.  Thus,
these index-based subgraph isomorphism methods do not enhance RDF
graph processing since there is only one big graph in an RDF database.


\section{Experiments}
\label{sec:experiment}

We perform extensive experiments on large-scale real and synthetic
datasets in order to show the superiority of a tamed subgraph
isomorphism algorithm for RDF query processing.  In the experiment, we
use \TurboISOP{}. We assume that \TurboISOp{} uses direct
transformation, while \TurboISOP{} uses type-aware transformation
along with all optimizations. The specific goals of the experiments
are 1) We show the superior performance of \TurboISOP{} over the
state-of-the-art RDF engines (Section~\ref{sec:comparison-other-rdf}),
2) We analyze the effect of the type-aware transformation and the
series of optimizations (Section~\ref{sec:effect-optimization}), and
3) We show the linear speed-up of the parallel \TurboISOP{} with an
increasing number of threads
\iftr
(Section~\ref{sec:effect-parall}).
\else
(Due to the space limit, please refer \cite{tr} for the detailed result).
\fi


\subsection{Experiment Setup}
\label{sec:experiment-setup}

\textbf{Competitors.} We choose three representative RDF engines as
competitors of \TurboISOP{} -- RDF-3X, TripleBit, and System-X. Note that
these three systems are publicly available. RDF-3X \cite{NW:10aa} is a
well-known RDF store, showing good performance for various types of
SPARQL queries. TripleBit \cite{YLW:13aa} is a very recent RDF engine
efficiently handling large-scale RDF data.
System-X is a popular RDF engine exploiting bitmap indexing.
We exclude BitMat \cite{ACZ:10aa} from performance evaluation since it is clearly inferior to TripleBit \cite{YLW:13aa}.
gStore is excluded since it is not publicly available.

\textbf{Datasets. }We use four RDF datasets in the experiment -- LUBM
\cite{GPH:05aa}, YAGO \cite{SKW:08aa},
BTC2012 \cite{H:12aa},
and BSBM \cite{BS:09aa}. LUBM
is a de-facto standard RDF benchmark which provides a
synthetic data generator. Using the generator, we create three
datasets -- LUBM80, LUBM800, and LUBM8000 where the number represents the
scaling factor.
YAGO is a real dataset which consists
of facts from Wikipedia and the WordNet.
BTC2012 is a real dataset crawled from multiple RDF web
resources. Lastly, BSBM is an RDF benchmark which
provides a synthetic data generator and benchmark queries. BSBM uses
more general SPARQL query features such as FILTER, OPTIONAL, and
UNION.
\iftr \else Due to the space limit, please refer \cite{tr} for the
experimental results for YAGO since the performance trends of YAGO are
similar to those for BTC2012.  \fi

In order to support the original benchmark queries in LUBM, we load
the original triples as well as inferred triples into databases.  In
order to obtain inferred triples, we use the state-of-the-art RDF
inference engine.  For example, LUBM8000 contains 1068394687 original
triples and 869030729 inferred triples. Note that this is the
\emph{standard} way to perform the LUBM benchmark.
However, regarding
BTC2012, we use the original triples only for database loading.  This
is because the BTC2012 dataset contains many triples that violate the
RDF standard, and thus the RDF inference engine refuses to load and
execute inference for the BTC2012 dataset.  BSBM contains 986410726
original triples and 11412064 inferred triples.

\edit{Table~\ref{tab:stat} shows the number of vertices and edges of
  the graphs transformed by the direct transformation and the
  type-ware transformation. The reduced number of edges in the
  type-aware transformed graph directly affects the amount of graph
  exploration in e-graph homomorphism matching.}

\begin{table}[h!]
  \centering
  \caption{\edit{Graph size statistics (direct: direct transformation, type-aware: type-aware transformation).}}
  \label{tab:stat}
  \resizebox{\linewidth}{!}{
  \begin{tabular}[h!]{|l|r|r|r|r|}
    \hline
    & $|V|$ direct & $|E|$ direct
    & $|V|$ type-aware & $|E|$ type-aware \\
    \hline
    LUBM80 & 2644579 & 19461754 & 2644573 & 12357312 \\
    LUBM800 & 26304872 & 193691328 & 26304863 & 122994224 \\
    LUBM8000 & 263133301 & 1937425416 & 263133295 & 1230263406 \\
    BTC2012 & 367728453 & 1436545556 & 367459811 & 1185887764 \\
    BSBM & 223938701 & 997822791 & 1937425416 & 893575906\\
    \hline
  \end{tabular}
  }
\end{table}

\begin{table*}[t]
  \centering
  \caption{Number of solutions in LUBM \edit{queries}.}
  \label{tab:lubm-sol}
  \setlength{\tabcolsep}{1.8mm}
  \resizebox{0.9\linewidth}{!}{
   \small
  \begin{tabular}{|c|r|r|r|r|r|r|r|r|r|r|r|r|r|r|r|r|r|r|}
    \hline
    Dataset & Q1 & Q2 & Q3 & Q4 & Q5 & Q6 & Q7
    & Q8 & Q9 & Q10 & Q11 & Q12 & Q13 & Q14 \\
    \hline
    LUBM80 & 4 & 212 & 6 & 34 & 719 & 838892 & 67 & 7790 & 21872 & 4 & 224 & 15 & 380 & 636529 \\
    LUBM800 & 4 & 2003 & 6 & 34 & 719 & 8352839 & 67 & 7790 & 218261 &
                                                                       4 & 224 & 15 & 3800 & 6336816 \\
    LUBM8000 & 4 & 2528 & 6 & 34 & 719 & 83557706 & 67 & 7790 & 2178420 & 4 & 224 & 15 & 37118 & 63400587 \\
    \hline
  \end{tabular}
  }
\end{table*}

\begin{table*}[t]
  \centering
  \caption{Elapsed time in LUBM [unit: ms] (X: wrong number of
    solutions (\# of solutions difference $>$ 3)
    , `*': wrong number of solutions (\# of solutions difference
    $\leq$ 3)).
  }
  \iftr \else \vspace{-3mm} \fi
  \label{tab:lubm-elap}
  \subfloat[LUBM80.]
  {
    \setlength{\tabcolsep}{1.5mm}
    \resizebox{0.9\linewidth}{!}{
    \begin{tabular}{|c|r|r|r|r|r|r|r|r|r|r|r|r|r|r|r|r|r|r|}
      \hline
      & Q1 & Q2 & Q3 & Q4 & Q5 & Q6 & Q7
      & Q8 & Q9 & Q10 & Q11 & Q12 & Q13 & Q14 \\
      \hline
      \TurboISOP{} & 0.09 & 6.37 & 0.09 & 0.13 & 0.13 & 4.43 & 0.05 & 2.26 & 101.42 & 0.09 & 0.10 & 0.10 & 0.06 & 3.08 \\
      RDF-3X & 3.09 & 188.90 & 4.09 & 12.37 & 14.74 & 375.04 & 91.06 & 58.32 & 770.32 & 3.19 & 2.35 & 3.52 & 15.08 & 262.41 \\
      TripleBit & 2.56 & 86.09 & 12.82 & 5.26 & 18.92 & 165.93 & 24.76 & 48.22$^*$ & X & 9.23 & 0.44 & 1.86 & 19.31 & 132.09 \\
      System-X & 2.00 & 426.00 & 2.00 & 4.67 & 2.67 & 64.33 & 4.00 & 19.33 & 3512.00 & 2.00 & 2.33 & 4.67 & 5.67 & 47.00 \\
      \hline
    \end{tabular}
    }
  }
  \\
  \subfloat[LUBM800.]
  {
    \setlength{\tabcolsep}{1.1mm}
    \resizebox{0.9\linewidth}{!}{
    \begin{tabular}{|c|r|r|r|r|r|r|r|r|r|r|r|r|r|r|r|r|r|r|}
      \hline
       & Q1 & Q2 & Q3 & Q4 & Q5 & Q6 & Q7
      & Q8 & Q9 & Q10 & Q11 & Q12 & Q13 & Q14 \\
      \hline
      \TurboISOP{} & 0.09 & 124.13 & 0.09 & 0.13 & 0.13 & 25.70 & 0.05 & 2.32 & 1239.46 & 0.09 & 0.10 & 0.09 & 0.12 & 19.72 \\
      RDF-3X & 4.15 & 2473.01 & 5.17 & 16.50 & 25.02 & 5103.35 & 840.16 & 461.80 & 10033.57 & 3.83 & 7.48 & 7.09 & 100.16 & 3607.13 \\
      TripleBit & 23.32 & 3548.58$^*$ & 142.29 & 15.76 & 183.46 & 2309.57 & 187.39 & 181.20$^*$ & X & 109.47 & 2.84 & 3.51 & 161.65 & 1818.52 \\
      System-X & 2.67 & 4394.00 & 2.00 & 4.67 & 3.00 & 239.33 & 4.33 & 21.00 & 175040.33 & 2.00 & 2.33 & 4.00 & 29.00 & 186.33 \\
      \hline
    \end{tabular}
    }
  }
  \\
  \subfloat[LUBM8000.]
  {
    \setlength{\tabcolsep}{0.7mm}
    \resizebox{0.9\linewidth}{!}{
      \small
    \begin{tabular}{|c|r|r|r|r|r|r|r|r|r|r|r|r|r|r|r|r|r|r|}
      \hline
       & Q1 & Q2 & Q3 & Q4 & Q5 & Q6 & Q7
      & Q8 & Q9 & Q10 & Q11 & Q12 & Q13 & Q14 \\
      \hline
      \TurboISOP{} & 0.10 & 309.74 & 0.09 & 0.12 & 0.13 & 191.52 & 0.05 & 1.61 & 5238.79 & 0.09 & 0.11 & 0.10 & 0.83 & 149.53 \\
      RDF-3X & 4.31 & 30492.93 & 4.87 & 19.53 & 94.89 & 65453.67 & 8476.19 & 4201.81 & 131053.33 & 4.15 & 23.27 & 12.83 & 630.91 & 48285.17 \\
      TripleBit & X & X & X & X & 2348.87 & 18974.80 & X & X & X & 1251.25 & X & X & X & 14197.47 \\
      System-X & 2.67 & 41449.33 & 2.67 & 5.00 & 3.00 & 1519.67 & 4.33 & 42.67 & 3123629.67 & 2.67 & 2.33 & 5.00 & 88.00 & 1155.00 \\
      \hline
    \end{tabular}
    }
  }
\vspace{-3mm}
\end{table*}

\textbf{Queries. }Regarding LUBM, we use the $14$ original benchmark queries
provided in the
website\footnote{\url{http://swat.cse.lehigh.edu/projects/lubm/}}. Previous work such as \cite{YLW:13aa} and
\cite{ZYW:13aa} modified some of the original queries
because executing those original queries without the inferred
triples returns an empty result set. Regarding
\iftr YAGO and \fi BTC2012, we use the same query sets proposed in
\iftr \cite{NW:10aa} and \fi
\cite{YLW:13aa},
because they do not have official benchmark
queries.
\iftr Some queries in the YAGO query set contain predicates
which do not exist in the YAGO dataset. We replace such predicates in
queries with the predicates in the dataset that have the closest
meaning. For example, the predicate \verb|bornInLocation| in Q1, Q5,
and Q6 is replaced with \verb|bornIn|.  \fi Regarding BSBM, we used 12
queries in the explore use case
\footnote{\url{http://wifo5-03.informatik.uni-mannheim.de/bizer/berlinsparqlbenchmark/spec/ExploreUseCase/index.html}}
which contain OPTIONAL, FILTER, and UNION keywords which test the
capability of more general SPARQL query support.

In order to measure the pure subgraph matching performance, (1) we
omit modifiers which reorganize the subgraph pattern matching results
(e.g. DISTINCT and ORDER BY) in all queries and (2) we measure the
elapsed time excluding the dictionary look-up time.

\textbf{Running Environment. }We conduct the experiments in a server running Linux four
Intel Xeon E5-4640 CPUs and $1.5TB$ RAM. The server has the NUMA
\cite{LPM:13aa,LBK:14aa} architecture with $4$ sockets in which each
socket has its own CPU and local memory.

We measure the elapsed times with a warm cache. To do that, we set up
the competitors' running environment as follows. For RDF-3X and
TripleBit, as done in \cite{ZYW:13aa}, we put the database files in
the \textit{tmpfs} in-memory filesystem, which is a kind of RAM disk.
For System-X, we set the memory buffer size to $400GB$, which is
sufficient for loading the entire database in memory. We execute every
query five times, exclude the best and worst times, and compute the
average of the remaining three.

\subsection{Comparison between Turbo$_{HOM}$++ and RDF
  engines}
\label{sec:comparison-other-rdf}

We report the elapsed times of the benchmark queries using a single thread.
Since the server has a NUMA architecture, memory allocation is always done within one CPU's local memory.

\textbf{LUBM. }Table~\ref{tab:lubm-sol} shows the number of solutions for all
benchmark queries in all LUBM datasets. Table~\ref{tab:lubm-elap}
shows experimental results for LUBM80, LUBM800, and LUBM8000. Note that Triplebit
was not able to return correct answers for two queries over
LUBM80/LUBM800 and for ten queries over LUBM8000.
In Table~\ref{tab:lubm-elap}, we use 'X' or the superscript `*' over
the elapsed times when TripleBit returns incorrect numbers of
solutions.

In order to analyze results in depth, we classify the LUBM queries into two types.
The first type of queries has a constant number of solutions regardless of the
dataset size. Q1, Q3 \verb|~| Q5, Q7, Q8, and Q10 \verb|~| Q12 belong to this type.
These queries are called \emph{constant solution queries}. The other queries (Q2,
Q6, Q9, Q13, and Q14) have increasing numbers of solutions proportional to the
dataset size. These queries are called \emph{increasing solution queries}.

Regarding the constant solution queries, only \TurboISOP{} achieves the ideal performance in LUBM, which means
constant performance regardless of dataset size. This phenomenon is analyzed as follows.
Each constant solution query contains a query vertex whose ID attribute is set to an entity in the RDF graph.
Thus, \TurboISOP{} chooses that query vertex as a starting query vertex and generates a candidate region.
Furthermore, in the LUBM datasets, although we increase the scaling factor in order to increase the database size,
the size of the candidate region explored by every constant solution query remains almost the same.

In contrast, the elapsed times of RDF-3X increase as the dataset size
increases. This is because the data size to scan for merge join
increases as the dataset size increases. Thus, the performance gap
between \TurboISOP{} and RDF-3X increases as the dataset size
increases.
In LUBM80, \TurboISOP{} is $23.50$ (Q11) \verb|~| $1821.20$
(Q7) times faster than RDF-3X. In LUBM800, \TurboISOP{} outperforms
RDF-3X by $42.56$ (Q10) \verb|~| $16803.20$ (Q7) times.
In LUBM8000,
\editt{\TurboISOP{} outperforms RDF-3X by $43.10$(Q1)} \verb|~| \editt{$169523.80$
(Q7) times. TripleBit shows a similar trend as RDF-3X.}
Accordingly,\TurboISOP{} is $4.40$ (Q11 in LUBM80) \verb|~| $18068.23$ (Q5 in
LUBM8000) times faster than TripleBit.
System-X shows constant elapsed
times for these queries, although it is consistently slower than
\TurboISOP{} by up to $86.60$ times.

For the increasing solution queries (Q2, Q6, Q9, Q13, and Q14),
\TurboISOP{} also shows the best performance in all LUBM datasets.
Overall, the elapsed times of \TurboISOP{} are
proportional to the number of solutions for these queries.
Specifically, after type-aware transformation, Q13 has one query vertex whose ID attribute is set to an entity in the data graph.
Thus, the number of candidate regions is one, which is similar to the constant solution query.
However, as the dataset size increases, the candidate region size also increases.
The other queries (Q2, Q6, Q9, Q14) do not have any query vertex whose ID attribute is set to an entity in the data graph.
As the dataset increases, the number of candidate regions for these queries increases, while each candidate region size does not
change.
All systems show the increasing elapsed time as the dataset size
increases.
RDF-3X shows $7.60$ (Q9 in LUBM80)
\verb|~| $760.13$ (Q13 in LUBM8000) times longer elapsed times than
\TurboISOP{}. TripleBit shows $13.51$ (Q2 in LUBM80) \verb|~|
$1347.08$ (Q13 in LUBM800) times longer elapsed time than \TurboISOP{}
when considering the queries which have the right number of solutions. System-X shows $7.72$ (Q14 in LUBM8000) \verb|~| $596.25$ (Q9 in
LUBM8000) times longer elapsed time than \TurboISOP{}.
For the constant solution query, System-X seems to be the best competitor of \TurboISOP{}. However, regarding the most time-consuming queries (Q2, Q9), System-X shows poor performance.

\iftr
\textbf{YAGO. } Since the YAGO dataset contains only about 50 million triples, all engines process the queries very efficiently.
Unlike the LUBM queries, the YAGO queries have only a few variables which are set to types.
Nevertheless, \TurboISOP{} exhibits the best performance for all YAGO queries.
Table~\ref{tab:yago} shows the exact number of solutions and elapsed
times in YAGO.

Specifically, \TurboISOP{} outperforms RDF-3X and System-X by up to
$25.95$ and $15.01$ times. This performance improvement is due to good
matching order selection and the series of optimizations in the
optimized \TurboISOP{}. Again, TripleBit returns incorrect numbers of
solutions for all queries except Q2.

\begin{table}[h!]
  \centering
  \caption{Number of solutions and elapsed time [unit: ms] in
    YAGO.}
  \iftr \else \vspace{-3mm} \fi
  \label{tab:yago}
\resizebox{\linewidth}{!}
{
  \setlength{\tabcolsep}{0.2mm}
  \begin{tabular}[t]{|c|r|r|r|r|r|r|r|r|}
    \hline
    & Q1 & Q2 & Q3 & Q4 & Q5 & Q6 & Q7 & Q8 \\
    \hline
    \# of sol. & 196 & 0 & 2129 & 3150 & 12611 & 2006 & 43238 & 91 \\
    \hline
    \TurboISOP{} & 1.33 & 0.13 & 16.93 & 1.19 & 3.61 & 20.52 & 31.35 & 4.04 \\
    RDF3X & 18.91 & 32.85 & 66.74 & 52.15 & 45.17 & 24.64 & 595.02 & 16.78 \\
    TripleBit & X & 1.03 & X & X & X & X & X & X \\
    System-X & 11.33 & 19.00 & 39.33 & 13.33 & 11.33 & 95.33 & 780.67 & 79.00 \\
    \hline
  \end{tabular}
}
\end{table}
\fi
\textbf{BTC2012. }Table~\ref{tab:btc2012} shows the exact number of solutions and
elapsed times in BTC2012. Even though BTC2012 contains over
$1$-billion triples, all the engines process all BTC2012 queries quite
efficiently. This is because the shapes of query graphs are simple
(tree-shaped). Furthermore, like LUBM, Q2, Q4, and Q5 in the BTC2012
query set contain one query vertex whose ID attribute is set to an
entity in the RDF graph. Still, \TurboISOP{} outperforms RDF-3X,
TripleBit, and System-X by up to 422.60, 28.57, and 266.18 times,
respectively.

\begin{table}[h!]
  \centering
  \caption{{Number of solutions and elapsed time [unit: ms] in BTC2012.}}
  \iftr \else \vspace{-3mm} \fi
  \label{tab:btc2012}
\resizebox{\linewidth}{!}
{
  \setlength{\tabcolsep}{0.2mm}
  \small
  \begin{tabular}[t]{|c|r|r|r|r|r|r|r|r|}
    \hline
    & Q1 & Q2 & Q3 & Q4 & Q5 & Q6 & Q7 & Q8 \\
    \hline
    \# of sol. & 4 & 4 & 1 & 4 & 13 & 1 & 664 & 5996 \\
    \hline
    \TurboISOP{} & 0.12 & 0.16 & 0.96 & 0.89 & 0.18 & 2.49 & 36.81 & 1.99 \\
    RDF3X & 6.67 & 7.52 & 10.42 & 13.07 & 69.97 & 22.75 & 392.73 & 841.96 \\
    TripleBit & 1.56 & 1.81$^*$ & 0.98 & 6.94 & 5.20 & 3.52 & 133.64$^*$ & X \\
    System-X & 8.00 & 4.67 & 5.00 & 12.33 & 4.67 & 663.67 & 110.67 & 351.67 \\
    \hline
  \end{tabular}
}
\end{table}

\textbf{BSBM.}
Table~\ref{tab:bsbm} shows the exact number of solutions and elapsed
times in BSBM. The open source RDF engines, RDF-3X and TripleBit, are
excluded as they do not support OPTIONAL and FILTER. Like BTC2012,
even though BSBM contains about $1$-billion triples, \TurboISOP{}
processes most BSBM queries less than 5ms except Q5 and Q6. That is
because they have a small number of solutions and contain one query
vertex whose ID attribute is set to an entity in the RDF graph. For
those ten queries, \TurboISOP{} outperforms System-X by 2.37 \verb|~|
7284.47 times. Q5 and Q6 take longer than the other queries because
they use expensive filters such as join conditions (Q5) and a
  regular expression (Q6) and filter out a large number of solutions
after basic graph pattern matching is finished. Before evaluating
FILTER, Q5 (Q6) has 178030 (2848000) solutions from the query graph
pattern and only qualifies 6803 (43508) final solutions.

\begin{table}[h!]
  \centering
  \caption{{Number of solutions and elapsed time [unit: ms] in BSBM.}}
  \iftr \else \vspace{-3mm} \fi
  \label{tab:bsbm}
  \setlength{\tabcolsep}{0.2mm}
  \begin{tabular}[t]{|c|r|r|r|r|r|r|}
    \hline
    & Q1 & Q2 & Q3 & Q4 & Q5 & Q6 \\
    \hline
    \# of sol. & 79 & 17 & 202 & 142 & 6803 & 43508 \\
    \hline
    \TurboISOP{} & 0.58 & 0.15 & 8.15 & 1.27 & 344.66 & 3969.18 \\
    System-X & 10 & 1092.67 & 19.33 & 21.67 & 589.67 & 9889.00  \\
    \hline
    & Q7 & Q8 & Q9 & Q10 & Q11 & Q12 \\
    \hline
    \# of sol. & 2 & 1 & 21 & 3 & 10 & 1 \\
    \hline
    \TurboISOP{} & 0.25 & 0.16 & 0.11 & 0.23 & 0.14 & 0.12\\
    System-X & 23.33 & 12.33 & 4.00& 11.00 & 3.00 & 8.00 \\
    \hline
  \end{tabular}
\end{table}

\setlength{\tabcolsep}{1.0mm}
\begin{table*}[t]
  \centering
  \caption{{Effect of type-aware transformation in LUBM8000 (Performance gain $=$ Direct transformation $\div$ Type-aware transformation).}}
  \iftr \else \vspace{-3mm} \fi
  \label{tab:effect-type-aware}
  \resizebox{\linewidth}{!}
    {
      \small
  \begin{tabular}[t]{|c|r|r|r|r|r|r|r|r|r|r|r|r|r|r|}
    \hline
    & Q1 & Q2 & Q3 & Q4 & Q5 & Q6 & Q7 & Q8 & Q9 & Q10 & Q11 & Q12 & Q13 & Q14 \\
    \hline
    Direct transformation (ms) & 0.101 & 57966.93 & 0.11 & 0.16 & 0.43 & 5218.47 & 0.15 & 5.63 & 114116.33 & 0.10 & 0.21 & 0.30 & 21.48 & 3886.43 \\
    Type-aware transformation (ms) & 0.100 & 50016.13 & 0.09 & 0.14 & 0.13 & 191.69 & 0.05 & 1.73 & 17829.50 & 0.09 & 0.11 & 0.10 & 1.33 & 149.60 \\
    Performance gain & 1.01 & 1.16 & 1.23 & 1.09 & 3.34 & 27.22 & 2.80 & 3.25 & 6.40 & 1.14 & 1.95 & 3.01 & 16.17 & 25.98 \\
    \hline
  \end{tabular}
    }
\end{table*}

\subsection{Effect of Improvement Techniques}
\label{sec:effect-optimization}

We measure the effect of the improvement
techniques including the type-aware transformation
(Section~\ref{sec:type-awre-transf}) and the four optimizations
(Section~\ref{sec:optimization}). For this purpose, we use the largest
LUBM dataset, LUBM8000. We first show the effect of the type-aware
transformation because it is beneficial to all LUBM queries. We next
show the effect of the four optimizations
(Section~\ref{sec:effect-five-optim}).


\subsubsection{Effect of Type-aware Transformation}
\label{sec:effect-type-aware}

Table~\ref{tab:effect-type-aware} shows the elapsed times for the LUBM
  queries in LUBM\-8000 using the direct transformation (\TurboISOp{})
  and the type-aware transformation (\TurboISOP{} without
  optimizations). Compared with the direct transformation, the
type-aware transformation improves the query performance by $1.01$(Q1)
to $27.22$(Q6).

The obvious reason for performance improvement is the smaller query
sizes after the type-aware transformation.
The reduced sized query graph
leads to smaller size candidate regions and shorter elapsed times.
First of all, Q6 and Q14 benefit the most from the type-aware
transformation. After the type-aware transformation, these queries
become point-shaped. That is, solutions of these two queries are
directly obtained by iterating the data vertices which have the vertex
label of the query vertex, which corresponds to lines 2--4 in
Algorithm~\ref{alg:turboiso}. 
Q13 also benefits much from the type-aware
transformation, since the type-aware transformation chooses a better starting query
vertex than the direct transformation which chooses a query vertex having type information.
Q1, Q3, Q4, Q5, Q7, Q8, Q10, Q11, and Q12 do not benefit from the
type-aware transformation because they already have a small number of
candidate vertices under the direct transformation.

\editt{Q2 benefits less than the other long running queries from the
  type-aware transformation. The following is the profiling result of
  Q2 with the direct/type-aware transformation. Q2 with direct
  transformation takes 26774.73 milliseconds in
  $ExploreCandidateRegion$ and 31191.29 milliseconds in
  $SubgraphSearch$.  Note that, with direct transformation, the
  starting vertex is arbitrarily chosen from $u_0$, $u_1$, $u_2$ in
  Figure~\ref{fig:direct-trans-query-graph-q2} since they all have
  same vertex label frequency ($freq(g,L(u_i)) = 1, i = 0,1,2$) and
  the same degree of $1$. In our implementation, the first query vertex
  $u_0$ is chosen and thus the label of the non-tree edge is
  \textsf{subOrganizationOf}. However, with type-aware transformation,
  the starting vertex is $u_1$ in
  Figure~\ref{fig:type-ware-trans-query-q2}, and the label of the
  non-tree edge is \textsf{memberOf}. Although the number of candidate
  regions with $u_1$ is the minimum among $u_0$, $u_1$, and $u_2$, the
  cost of $IsJoinable$ calls for \textsf{memberOf} increases 1.30
  times. Thus, Q2 with type-aware transformation takes 9523.60
  milliseconds in $ExploreCandiateRegion$ and 40469.47 milliseconds in
  $SubgraphSearch$. We achieve only 1.16 times performance
  improvement. However, the cost of the $IsJoinable$ call is significantly
  reduced by using \textsf{+INT}. Thus, after applying type-aware
  transformation and the tailored optimizations, the final elapsed
  time for Q2 becomes 309.74ms, i.e., 187.15 times performance
  improvement compared with direct transformation only.}



\subsubsection{Effect of Four Optimizations}
\label{sec:effect-five-optim}

In this experiment, we measure the effect of four optimizations of
\TurboISOP{}. We use Q2 and Q9 in LUBM8000 since these two
queries in LUBM8000 are the most time-consuming and exploit all
optimizations. All the other queries are omitted since
  their elapsed times are too short, so that it is hard to recognize the effect
  of optimization. Note that the elapsed times of Q1, Q3 \verb|~|
Q5, Q7, Q8, Q10 \verb|~| Q13 are too short ($< 2ms$), and
  Q6 and Q14 do not benefit from these optimizations since they are
  point-shaped.

  Figure~\ref{fig:effect-opt} shows \edit{the reduced times} of Q2 and Q9 in
  LUBM8000 after applying these optimizations \edit{separately}. The
  optimization techniques in X-axis are ordered by the reduced in a decreasing
  manner --- \textsf{+INT}, \textsf{-NLF}, \textsf{-DEG}, and \textsf{+REUSE}.
  \edit{Interestingly, even though Q2 and Q9 have the same shape
    (i.e., trianglular), the most effective optimizations were
    different. \textsf{+INT} was the most effective in
    Q2. \textsf{-NLF} was the most effective in Q9 since the size of
    each candidate region was very small. \textsf{-DEG} was more
    effective in Q9 than in Q2 since Q9 has more data vertices applied
    to the degree filter. \textsf{+REUSE} was effective in Q9 which
    has large number of candidate regions while Q2 did not benefit
    from \textsf{+REUSE}. \\}


\begin{figure}[h!]
  \centering
  \includegraphics[width=0.9\linewidth]{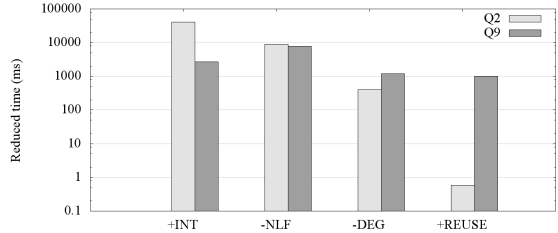}
  \caption{{Reduced elapsed time of each optimization \edit{(Elapsed time of no-optimization:
    50016.13ms (Q2) and 17829.50ms (Q9))}.}}
  \label{fig:effect-opt}
\end{figure}

\iftr
\subsection{Effect of Parallelization}
\label{sec:effect-parall}

In the last experiment, we report the parallelization effect of
\TurboISOP{}. Among parallelizable queries (Q2, Q6, Q9, and
  Q14), which have multiple starting data vertices, we choose Q2 and
  Q9. The reasons are 1) these queries are the most time-consuming
  queries, and 2) Q6 and Q14 are point-shaped queries which do not
  involves graph exploration. In order to show the parallelism, we
allocate the data graph in an interleaved way where each memory page
allocation is assigned to sockets in a round-robin way.

We vary the number of threads by $1$, $4$, $8$, $12$ and $16$.
As shown in
  Figure~16,
\TurboISOP{} shows super-linear speed-up proportional to
the number of threads. In Q2, \TurboISOP{} achieves $5.37$, $10.49$,
$15.06$ and $19.63$ speed-up using $4$, $8$, $12$, and $16$ threads,
respectively. In Q9, \TurboISOP{} achieves $4.87$, $8.23$, $14.37$ and
$16.54$ speed-up. In the experiment, though the data graph is
evenly spread out in 4 sockets, data vertices of a candidate region could
not be uniformly distributed. That means executing a pattern matching of
the candidate region in a socket which has more data vertices in its local memory
for the candidate region is beneficial since less remote memory access is
required. In this sense, measuring speed-up based on a multiple of 4
threads is more reasonable. When computing the speedup based on the
4-thread elapsed time, the speed-up of $Q2$ becomes $1.95$, $2.80$,
and $3.65$ for $8$, $12$, and $16$ threads.


\begin{figure}[h!]  \label{fig:effect-parallel-fig}
  \centering
  \includegraphics[width=0.6\linewidth]{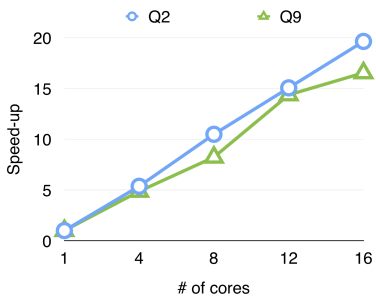}
  \caption{Speed-up of \TurboISOP{} in Q2 and Q9 in the LUBM8000 dataset).}
\end{figure}
\fi 

\vspace{-3mm}
\section{Conclusion}
\label{sec:conclusion}

The core function of processing RDF data is subgraph pattern matching. There have been two completely different directions for supporting efficient subgraph pattern matching. One direction is to develop specialized RDF query processing engines exploiting the properties of RDF data, while the other direction is to develop efficient subgraph isomorphism algorithms for general, labeled graphs. In this paper, we posed an important research question, ``\emph{Can subgraph isomorphism be tamed for efficient RDF processing?}''
In order to address this question, we provided the first direct and comprehensive comparison of the state-of-the-art subgraph isomorphism method with representative RDF processing engines.

We first showed that a subgraph isomorphism algorithm requires minimal
modification to handle a graph homomorphism with edge label mapping
which is the RDF graph pattern matching semantics. We then provided a
novel transformation method, called type-aware transformation along
with a series of optimization techniques. We next performed extensive
experiments using RDF benchmarks in order to show the superiority of
the optimized subgraph isomorphism over representative RDF processing
engines.  Experimental results showed that the optimized subgraph
isomorphism method achieved consistent and significant speedup over
those RDF processing engines.

This study drew a promising conclusion that a
  subgraph isomorphism algorithm tamed for RDF processing can serve as an in-memory accelerator on top of a commercial RDF engine for \emph{real-time}
  RDF query processing as well. We believe that this approach opens a
  new direction for RDF processing, so that both traditional
  directions can merge or benefit from each other.



\section*{Acknowledgment}

This work was supported in part by a gift from Oracle Labs' External Research Office. This work was also
supported by the National Research Foundation of Korea(NRF)
grant funded by the Korea government(MSIP) (No. NRF-2014R1A2A2A01004454) and
the MSIP(Ministry of Science, ICT and Future Planning),
Korea, under the ``ICT Consilience Creative Program'' (IITP-2015-R0346-15-1007)
supervised by the IITP(Institute for Information \&  communications Technology Promotion).

\balance
\bibliographystyle{abbrvnatsimple}
\bibliography{turbordf}  

\begin{thebibliography}{35}
\providecommand{\natexlab}[1]{#1}
\providecommand{\url}[1]{\texttt{#1}}
\expandafter\ifx\csname urlstyle\endcsname\relax
  \providecommand{\doi}[1]{doi: #1}\else
  \providecommand{\doi}{doi: \begingroup \urlstyle{rm}\Url}\fi

\bibitem[Abadi et~al.(2009)]{AMM:09aa}
D.~J. Abadi et~al.
\newblock Sw-store: A vertically partitioned dbms for semantic web data
  management.
\newblock \emph{The VLDB Journal},  385--406, 2009.

\bibitem[Atre et~al.()]{ACZ:10aa}
M.~Atre et~al.
\newblock Matrix "bit" loaded: A scalable lightweight join query processor for
  rdf data.
\newblock In \emph{WWW '10},  41--50.

\bibitem[Bizer and Schultz(2009)]{BS:09aa}
C.~Bizer and A.~Schultz.
\newblock The berlin sparql benchmark.
\newblock \emph{International Journal on Semantic Web and Information Systems
  (IJSWIS)},  1--24, 2009.

\bibitem[Broekstra et~al.()]{BKH:02aa}
J.~Broekstra et~al.
\newblock Sesame: A generic architecture for storing and querying rdf and rdf
  schema.
\newblock In \emph{ISWC '02},  54--68.

\bibitem[Cheng et~al.()]{CKN:07aa}
J.~Cheng et~al.
\newblock Fg-index: Towards verification-free query processing on graph
  databases.
\newblock In \emph{SIGMOD '07},  857--872.

\bibitem[Fan et~al.()]{FLM:10aa}
W.~Fan et~al.
\newblock Graph homomorphism revisited for graph matching.
\newblock \emph{VLDB '10},  1161--1172.

\bibitem[Gubichev and Neumann()]{GN:14aa}
A.~Gubichev and T.~Neumann.
\newblock Exploiting the query structure for efficient join ordering in
  {SPARQL} queries.
\newblock In \emph{EDBT '14},  439--450.

\bibitem[Guo et~al.(2005)]{GPH:05aa}
Y.~Guo et~al.
\newblock Lubm: A benchmark for owl knowledge base systems.
\newblock \emph{Web Semant.},  158--182, 2005.

\bibitem[Han et~al.()]{HLL:13aa}
W.-S. Han et~al.
\newblock Turbo$_{ISO}$: towards ultrafast and robust subgraph isomorphism
  search in large graph databases.
\newblock In \emph{SIGMOD '13},  337--348.

\bibitem[Harth(2012)]{H:12aa}
A.~Harth.
\newblock {Billion Triples Challenge} data set.
\newblock Downloaded from http://km.aifb.kit.edu/projects/btc-2012/, 2012.

\bibitem[He and Singh()]{HS:08aa}
H.~He and A.~K. Singh.
\newblock Graphs-at-a-time: Query language and access methods for graph
  databases.
\newblock In \emph{SIGMOD '08},  405--418.

\bibitem[Huang et~al.()Huang, Abadi, and Ren]{HAR:11aa}
J.~Huang, D.~J. Abadi, and K.~Ren.
\newblock Scalable sparql querying of large rdf graphs.
\newblock \emph{VLDB '11},  1123--1134.

\bibitem[Lee et~al.()]{LHK:12aa}
J.~Lee et~al.
\newblock An in-depth comparison of subgraph isomorphism algorithms in graph
  databases.
\newblock \emph{VLDB '12},  133--144.

\bibitem[Leis et~al.()]{LBK:14aa}
V.~Leis et~al.
\newblock Morsel-driven parallelism: A numa-aware query evaluation framework
  for the many-core age.
\newblock In \emph{SIGMOD '14},  743--754.

\bibitem[Li et~al.(2013)Li, Pandis, M{\"u}ller, Raman, and Lohman]{LPM:13aa}
Y.~Li, I.~Pandis, R.~M{\"u}ller, V.~Raman, and G.~M. Lohman.
\newblock Numa-aware algorithms: the case of data shuffling.
\newblock In \emph{CIDR}, 2013.

\bibitem[Neumann and Moerkotte()]{NM:11aa}
T.~Neumann and G.~Moerkotte.
\newblock Characteristic sets: Accurate cardinality estimation for rdf queries
  with multiple joins.
\newblock In \emph{ICDE '11},  984 -- 994.

\bibitem[Neumann and Weikum()]{NW:10ab}
T.~Neumann and G.~Weikum.
\newblock x-rdf-3x: fast querying, high update rates, and consistency for rdf
  databases.
\newblock \emph{VLDB '10},  256--263.

\bibitem[Neumann and Weikum(2010)]{NW:10aa}
T.~Neumann and G.~Weikum.
\newblock The rdf-3x engine for scalable management of rdf data.
\newblock \emph{The VLDB Journal},  91--113, 2010.

\bibitem[P.~Cordella et~al.(2004)]{PFS:04aa}
L.~P.~Cordella et~al.
\newblock A (sub)graph isomorphism algorithm for matching large graphs.
\newblock \emph{IEEE Trans. Pattern Anal. Mach. Intell.},  1367 -- 1372, 2004.

\bibitem[Shang et~al.()]{SZL:08aa}
H.~Shang et~al.
\newblock Taming verification hardness: An efficient algorithm for testing
  subgraph isomorphism.
\newblock \emph{VLDB '08},  364--375.

\bibitem[Shao et~al.()]{SWL:13aa}
B.~Shao et~al.
\newblock Trinity: A distributed graph engine on a memory cloud.
\newblock In \emph{SIGMOD '13},  505--516.

\bibitem[Suchanek et~al.(2008)]{SKW:08aa}
F.~M. Suchanek et~al.
\newblock Yago: A large ontology from wikipedia and wordnet.
\newblock \emph{Web Semant.},  203--217, 2008.

\bibitem[Ullmann(1976)]{U:76aa}
J.~R. Ullmann.
\newblock An algorithm for subgraph isomorphism.
\newblock \emph{J. ACM},  31--42, 1976.

\bibitem[Weiss et~al.()]{WKB:08aa}
C.~Weiss et~al.
\newblock Hexastore: sextuple indexing for semantic web data management.
\newblock \emph{VLDB '08},  1008--1019.

\bibitem[Wilkinson and Wilkinson()]{WW:06aa}
K.~Wilkinson and K.~Wilkinson.
\newblock Jena property table implementation.
\newblock In \emph{SSWS '06},  35--46.

\bibitem[Yan et~al.()]{YYH:04aa}
X.~Yan et~al.
\newblock Graph indexing: A frequent structure-based approach.
\newblock In \emph{SIGMOD '04},  335--346.

\bibitem[Yan et~al.(2005)]{YYH:05aa}
X.~Yan et~al.
\newblock Graph indexing based on discriminative frequent structure analysis.
\newblock \emph{ACM Trans. Database Syst.},  960--993, 2005.

\bibitem[Yuan et~al.()]{YLW:13aa}
P.~Yuan et~al.
\newblock Triplebit: a fast and compact system for large scale rdf data.
\newblock \emph{VLDB '13},  517--528.

\bibitem[Zeng et~al.()]{ZYW:13aa}
K.~Zeng et~al.
\newblock A distributed graph engine for web scale rdf data.
\newblock \emph{VLDB '13},  265--276.

\bibitem[Zhang et~al.({\natexlab{a}})]{ZHY:07aa}
S.~Zhang et~al.
\newblock Treepi: A novel graph indexing method.
\newblock In \emph{ICDE '07},  966 -- 975, {\natexlab{a}}.

\bibitem[Zhang et~al.({\natexlab{b}})]{ZLY:09aa}
S.~Zhang et~al.
\newblock Gaddi: Distance index based subgraph matching in biological networks.
\newblock In \emph{EDBT '09},  192--203, {\natexlab{b}}.

\bibitem[Zhao and Han()]{ZH:10aa}
P.~Zhao and J.~Han.
\newblock On graph query optimization in large networks.
\newblock \emph{VLDB '10},  340--351.

\bibitem[Zou et~al.({\natexlab{a}})]{ZCY:08aa}
L.~Zou et~al.
\newblock A novel spectral coding in a large graph database.
\newblock In \emph{EDBT '08},  181--192, {\natexlab{a}}.

\bibitem[Zou et~al.({\natexlab{b}})]{ZMC:11aa}
L.~Zou et~al.
\newblock gstore: answering sparql queries via subgraph matching.
\newblock \emph{VLDB '11},  482--493, {\natexlab{b}}.

\end{thebibliography}

\end{document}